\documentclass[%
 reprint, 
 amsmath,amssymb,
 showpacs,
 aps, prb,
 longbibliography
]{revtex4-2}

\usepackage{graphicx}% Include figure files
\usepackage{dcolumn}% Align table columns on decimal point
\usepackage{bm}% bold math
\usepackage{hyperref}% add hypertext capabilities
\usepackage{multirow}% add support for complex tables
\hypersetup{
    colorlinks=true,
    linkcolor=blue,
    citecolor=blue,
    urlcolor=blue,
}% Added by user
\allowdisplaybreaks% Added by user

\begin{document}
\preprint{APS/123-QED}

\title{\textbf{Role of Ferrons in the Heat Capacity and Thermal Transport of Displacive Ferroelectrics} 
}% Force line breaks with \\

\author{G. D. Zhao}
 \affiliation{
 Department of Materials Science and Engineering, Pennsylvania State University, University Park, Pennsylvania 16802, USA
}

\author{F. Yang}
 \email{fzy5099@psu.edu}
 \affiliation{
 Department of Materials Science and Engineering, Pennsylvania State University, University Park, Pennsylvania 16802, USA
}

\author{L. Q. Chen}
 \email{lqc3@psu.edu}
 \affiliation{
 Department of Materials Science and Engineering, Pennsylvania State University, University Park, Pennsylvania 16802, USA
}

\date{\today}

\begin{abstract}
The collective amplitude mode of the order parameter in displacive ferroelectrics, termed the ferron, represents the amplitude fluctuations of long-range ordered polarization.
At temperatures well below phase transition temperature $T_c$, the energy of ferron excitation is significantly gapped in the long-wavelength limit. As $T_c$ is approached, this gap softens dramatically to minimal or gapless values, thereby should lead to a substantial contribution to thermal properties.
In this context, we explore the role of ferrons in heat capacity and thermal transport by incorporating a microscopic self-consistent phase-transition theory for displacive ferroelectricity in contrast to the conventional treatment of attributing thermal properties solely to acoustic phonons.
Using ferroelectric $\rm{PbTiO}_{3}$ as a case study, we show that the softening of ferrons near the phase transition is essential to accurately capturing the experimental temperature and electric-field dependencies of thermal properties.
\end{abstract}

\pacs{77.80.Bh, 05.30.-d, 77.80.Fm}
%77.80.Bh Phase transitions and Curie point
%05.30.−d Quantum statistical mechanics
%77.80.Fm Switching in ferroelectrics

\maketitle
%--------------------------------------------------------------------------
%                        Main Text
%--------------------------------------------------------------------------
%\tableofcontents

%Intro-1
\textit{Introduction.---}Ferroelectrics are known for their outstanding dielectric, pyroelectric, piezoelectric, and electrocaloric properties.
Beyond these electrical characteristics, their thermophysical properties---including specific heat, thermal diffusivity, and thermal conductivity---are critical performance metrics for applications such as thermal management for nanostructured microelectronics~\cite{Cahill2003,Bell2008,Shuai2017,Zou2024,Zhang2024}.
These thermal properties often limit the performance and reliability of state-of-the-art devices, emphasizing their importance in operational environments.
Despite this, most computational efforts aimed at understanding the thermal properties of ferroelectrics have focused on acoustic phonons, owning to their gapless energy spectrum.
While the phonon theory~\cite{Frenkel1932} has been instrumental to explaining the thermal behaviours, it often fails to accurately describe systems undergoing phase transitions, such as magnetic or ferroelectric materials.
For instance, the ferroelectric-paraelectric phase transition in displacive ferroelectrics~\cite{Levanyuk1974,Tani1969} is fundamentally distinct from processes dominated by acoustic phonons.
This transition is driven by the condensation of an unstable optical phonon mode at Brillouin zone center upon cooling~\cite{Cochran1959,Blinc1987}, which breaks the structural inversion symmetry. 
Experimentally, this transition is accompanied by unusual thermal behaviours~\cite{Yoshida1960,Remeika1970,Tachibana2022}: a rapid increase in specific heat and a slower-than-expected decrease in thermal conductivity over a wide temperature range leading up to the critical temperature $T_c$.
Such observations cannot be explained by the conventional Debye model, necessitating the inclusion of polarization-related phenomena to develop a consistent theoretical framework.

\begin{figure}[htbp]
\centering
\includegraphics[width=8.2cm]{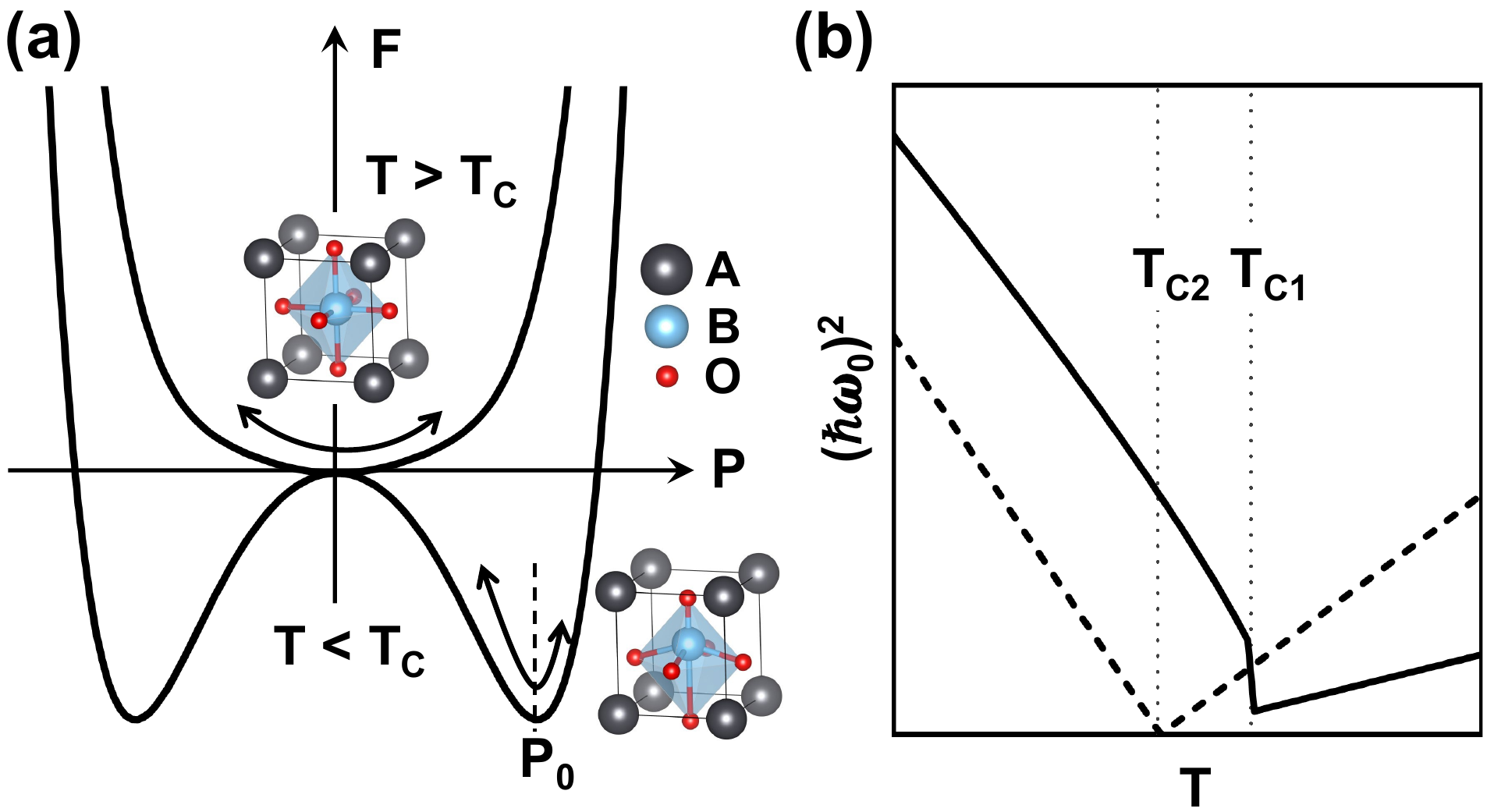}
    \caption{(a) Schematic illustration of polarization fluctuation in paraelectric phase above $T_c$ and in ferroelectric phase below $T_c$ in a ABO$_3$ perovskite material. 
    Arrows indicate the fluctuation of polarization. $P_0$ denotes the spontaneous polarization in the FE phase. 
    (b) A sketch of temperature-dependent squared quasi-particle gap $(\hbar\omega_{q=0})^2$
    across the first-order (solid curve) and second-order (dashed curve) FE-PE transitions.
    }
    \label{fig:fig1}
\end{figure}

%Intro-2
According to quantum statistical physics and Landau phase transition theory, the quasi-particles~\cite{Landau2013-5,Abrikosov2012} emerging in an ordered phase with spontaneous breaking of continuous symmetry play a pivotal role in determining macroscopic system properties. 
In ferromagnetic/antiferromagnetic materials, such excitations, termed magnons~\cite{Bloch1930,Holstein1940,Dyson1956,Kruglyak2010,Chumak2015}, are well-documented for their critical contributions to thermal properties. Their analogous excitations in ferroelectrics have received significantly less attention.
Recently, a quasi-particle excitation known as the “ferron” has been proposed in ferroelectrics~\cite{Bauer2021,Tang2022}, representing a natural analogue to the magnon. 
Ferrons describe the collective amplitude fluctuations of long-range ordered polarization in displacive ferroelectrics, as depicted schematically in Fig.~\ref{fig:fig1}(a).
This description offers a key quantum statistical mechanism to link polarization dynamics with macroscopic thermal properties, addressing a critical gap in existing theories.

%Intro-3
The ferron energy spectrum is derived from the free energy expression for the long-range ordered polarization $P(r)$, which is expressed as a sixth-order polynomial~\cite{Lines2001}:
 \begin{align}
    F_P\!=\!\!\!\int d^3 r \Big[
      \frac{\alpha(T)}{2} P^2 
    + \frac{\beta(T)}{4} P^4 
    + \frac{\lambda}{6} P^6
    + \frac{g}{2}(\nabla P)^2
        \Big],\label{eqn:free}
\end{align}
where $\alpha(T)$, $\beta(T)$, and $\lambda$ are Landau coefficients, $g$ is the gradient coefficient accounting for the energy cost of spatial inhomogeneity/fluctuation of the polarization.
Below the critical temperature, the equilibrium long-range ordered polarization $P_0^2$ orders in a first (second)-order phase transition for $\beta<0$ ($\beta>0$), and evolves as:
 \begin{equation}\label{eqn:P0}
    P_0^2=(-\beta+\sqrt{\beta^2-4\alpha\lambda})/(2\lambda).
 \end{equation}
The ferron energy spectrum  is then given by~\cite{Tang2022} 
\begin{align}
    \omega_q(T)=m_p^{-1/2} (\alpha + 3{\beta}P_0^2 + 5{\lambda}P_0^4 + gq^2)^{1/2},\label{eqn:wq}
\end{align}
where $m_p$ denotes the polarization inertia, or the effective mass of polarization~\cite{Sivasubramanian2004,Resta2003,Landau2013-8}.
Clearly, the excitation energy gap of ferron at the long-wavelength limit, $\omega_{q=0}(T)$, softens with increasing temperature, as shown in Fig.~\ref{fig:fig1}(b) (see in full temperature range in Appendix~\ref{app:wq}).
Around the criticality, for second-order phase transitions, $\hbar \omega_{q=0}\;(T{\sim}T_c)$ vanishes due to the continuous disappearance of the order parameter, implying a gapless ferron around the phase transition point. 
For first-order phase transitions, $\hbar\omega_{q=0}\;(T{\sim}T_c)$ approaches a finite yet minimal value compared to $k_B T_c$, reflecting the discontinuous nature of the order parameter and the typically high transition temperatures of ferroelectrics.
The softening feature is a fundamental property of quasi-particles in ordered phases: the long-wavelength gap is related to the order parameter, and  protects the existence of quasiparticle, causing the thermal anomalies near criticality.

%Intro-4
Due to the characteristic softening feature of the ferron excitation energy gap $\omega_{q=0}$ with increasing temperature, ferron thermal excitations are expected to play a critical role in the thermodynamic and transport properties of ferroelectrics, particularly near the phase transition temperature $T_c$, where the energy gap becomes minimal or vanishes.
In this regime, ferrons exhibit behaviour akin to acoustic phonons (see, e.g., Fig.~\ref{fig:figs1}), making their contributions to heat capacity and thermal transport comparable to those of acoustic phonons and, therefore, non-negligible in determining the overall thermal properties.

%Intro-5
In this work, we investigate the thermophysical properties of displacive ferroelectrics by incorporating contributions from both ferrons and acoustic phonons.
Specifically, in the quantum-statistical mechanism and thermal transport calculations for ferrons, we employ a self-consistent microscopic phase-transition theory to incorporate the temperature-dependent softening feature of the excitation energy gap $\omega_{q=0}$. 
For acoustic phonons, we utilize the Debye model~\cite{Kittel2005,Callaway2013} for their contribution, with their temperature dependence extracted from experimental sound velocity measurements.
Together, these two components account for the dominant low-energy excitations in ferroelectrics: as ferrons soften near phase transitions and acoustic phonons remain gapless, their statistical nature as bosons under the Bose-Einstein distribution ensures that they primarily govern the temperature dependence of thermal properties.
Then, using ferroelectric PbTiO$_3$ as a case study, we calculate the heat capacity and thermal conductivity, demonstrating that only by including the ferron contribution can the theoretical predictions quantitatively capture the experimentally observed temperature and external electric-field dependencies in thermal properties.
This work bridges the gap between classical acoustic phonon models and experimental anomalies in displacive ferroelectrics, providing a unified framework for understanding field-sensitive thermal transport properties across ferroelectric phase transitions.

%Section_1-1
\textit{Model.---}In order to self-consistently incorporate the temperature-dependent softening of the ferrons, we solve the excitation energy in Eq.~(\ref{eqn:wq}) alongside order-parameter equation in Eq.~(\ref{eqn:P0}). Particularly, the Landau coefficients for the order parameter are accurately determined by using the recently developed self-consistent microscopical phase-transition theory of displacive ferroelectricity~\cite{yang2024microscopic}. 
This approach allows the prediction of finite-temperature properties across the entire temperature range of the phases, particularly capturing the critical behaviour near the phase transition.

%Section_1-2
Using quantum statistic theory, the molar specific heat of ferrons (amplitude fluctuation of the long-range ordered polarization) at constant volume is derived as~\cite{Callaway2013}
(see derivations in Appendix~\ref{app:pt}):
\begin{equation}
    c_v
    \!=\! N_AV_0\frac{du}{dT}\!=\! 
    \frac{N_A V_0 \hbar}{2\pi^2}
    \int_0^{q_D}\!\! dq\;q^2
    \Big(
        \frac{\partial \omega_q}{\partial T}\bar{n}_q 
        \!+\! \omega_q \frac{\partial \bar{n}_q}{\partial T}
    \Big)
    ,\label{eqn:cv}
\end{equation}
where $u=\sum_{\bf q}\hbar\omega_{q}{\bar n}_{q}$ denotes the total energy per unit reciprocal volume of ferrons in terms of quantum statistic mechanics. 
Here, $V_0$ denotes the unit cell volume, $\hbar$ is the reduce Plank constant, ${N_A}$ is the Avogadro constant, and $\bar{n}_q=[\exp(\hbar\omega_q/(k_B T))-1]^{-1}$ represents the Bose-Einstein distribution at thermal equilibrium, with ${k_B}$ being the Boltzmann constant. 
It should be emphasized that unlike the conventional treatments within Debye model and standard Boltzmann transport equation with temperature-independent excitation energy, here we have explicitly retained $\partial \omega_q/\partial T$, which arises from the temperature-dependent energy softening of the ferrons. Conventional BTE approaches typically assume a fixed phonon dispersion and do not incorporate this essential derivative, making them less effective in describing strong temperature-dependent softening near phase transitions.
The isotropic integration over wavevector space up to a Debye-like cutoff $q_D$ is used solely to normalize the number of ferron modes, in analogy with Debye’s treatment of phonons. 
This approximation enables tractable analytical evaluation of ferron contributions and does not assume linear dispersion.

%Section_1-3
The thermal conductivity $\kappa$ due to ferrons is derived by using the semi-classical Boltzmann transportation equation (BTE), yielding (see derivations in Appendix~\ref{app:bte}):
\begin{eqnarray}
    \kappa
    &=&
    \frac{\tau g^2\hbar}{6\pi^2m_p^2}
    \int_0^{q_D}\!\!dq \; \frac{q^4}{\omega_q} \frac{d \bar{n}_q}{d T} \nonumber\\
    &=& 
    \frac{\tau g^2\hbar^2}{6\pi^2m_p^2 k_B T}
    \int_0^{q_D}\!\!dq \; \frac{q^4}{\omega_q}  
    \left(
    \frac{\omega_q}{T} - \frac{\partial \omega_q}{\partial T}
    \right)
    \bar{n}_q (\bar{n}_q + 1),~~~
    \label{eqn:kappa}
\end{eqnarray}
where, again, the temperature dependence of $\omega_q$ is explicitly considered to self-consistently account for ferron softening. This inclusion goes beyond conventional BTE treatments, which do not incorporate $\partial \omega_q/\partial T$ and thus may miss key contributions from strongly temperature-dependent quasiparticles near phase transitions.
%where, again, the temperature dependence of $\omega_q$, is explicitly considered to self-consistently account for ferron softening. 

%Section_1-4
The momentum scattering/relaxation time $\tau$ is primarily dictated by the third-order anharmonic interactions associated with the three-phonon processes. Taking account of the scattering probabilities of the phonon emission and absorption within the microscopic scattering mechanism~\cite{yang2015hole,yang2016spin,yang2025THz},  the scattering rate should be proportional to the acoustic phonon population, $\bar{n}_{ac,q}$, such that $\tau^{-1}\propto2\bar{n}_{ac,q} + 1$. 
At elevated temperatures, typically above the Debye temperature of acoustic phonons, the relation simplifies to $\tau^{-1}\propto{T}$.
Therefore, we approximate the relaxation time as $\tau(T)=\tau(T=T_{\rm RT})\times{T_{\rm RT}}/T$, where $T_{\rm RT}$ is the room temperature and $\tau(T=T_{\rm RT})$ is the relaxation time observed at room temperature. 
While assuming the relaxation time is an approximation, it remains valid within the linear response regime considered in this work. 
Besides, the relaxation time used in our calculations is chosen to be consistent with experimental values.
Therefore, the omission of fully microscopic and complete bosonic scattering treatment does not affect our main results and conclusions.

%Section_1-5
Further, by incorporating the temperature dependences of both $\alpha$ and $\beta$ in ferroelectric PbTiO$_3$, the essential energy softening contribution of ferrons to the specific heat and thermal conductivity can be expressed as

\begin{align}
    \frac{\partial \omega_q}{\partial T}
    \!=\!\frac{1}{2m_p\omega_q}
    \left[ 
    \frac{\partial \alpha}{\partial T}\!+\!3 P_0^2 \frac{\partial \beta}{\partial T}
    \!+\!(3\beta\!+\!10\lambda P_0^2)\frac{\partial P_0^2}{\partial T}
    \right]
    ,\label{eqn:pwpt}
\end{align}
which has been ignored in conventional treatments~\cite{Callaway2013,Li2014}, and the pyroelectric effect is accounted as:
\begin{align}%\small
    \frac{\partial P_0^2}{\partial T} 
    = \frac{1}{2\lambda}
    \left[
        \left(\beta \frac{\partial \beta}{\partial T} 
        \!-\!2\lambda \frac{\partial \alpha}{\partial T}\right) 
        \left(\beta^2\!-\! 4\alpha\lambda\right)^{-\frac{1}{2}}
    \!-\!\frac{\partial \beta}{\partial T} 
    \right]. \label{eqn:pps2pt}
\end{align}
For the acoustic phonon contributions to the specific heat and thermal conductivity, we employ the conventional Debye model, with detailed calculations provided in the Appendix~\ref{app:pt}. 

%section_2-1
\textit{Results of ferroelectric PbTiO$_3$.---}We consider the ferroelectric PbTiO$_3$ ~\cite{Burns1970,Burns1973,Shirane1970,Cohen1992,Rabe2007} as a specific example, where a first-order ferroelectric-paraelectric phase transition occurs at $T_c\approx765\;$K between its high-temperature cubic phase and low-temperature tetragonal phase.

\begin{figure}[htbp]
\centering
\includegraphics[width=8.2cm]{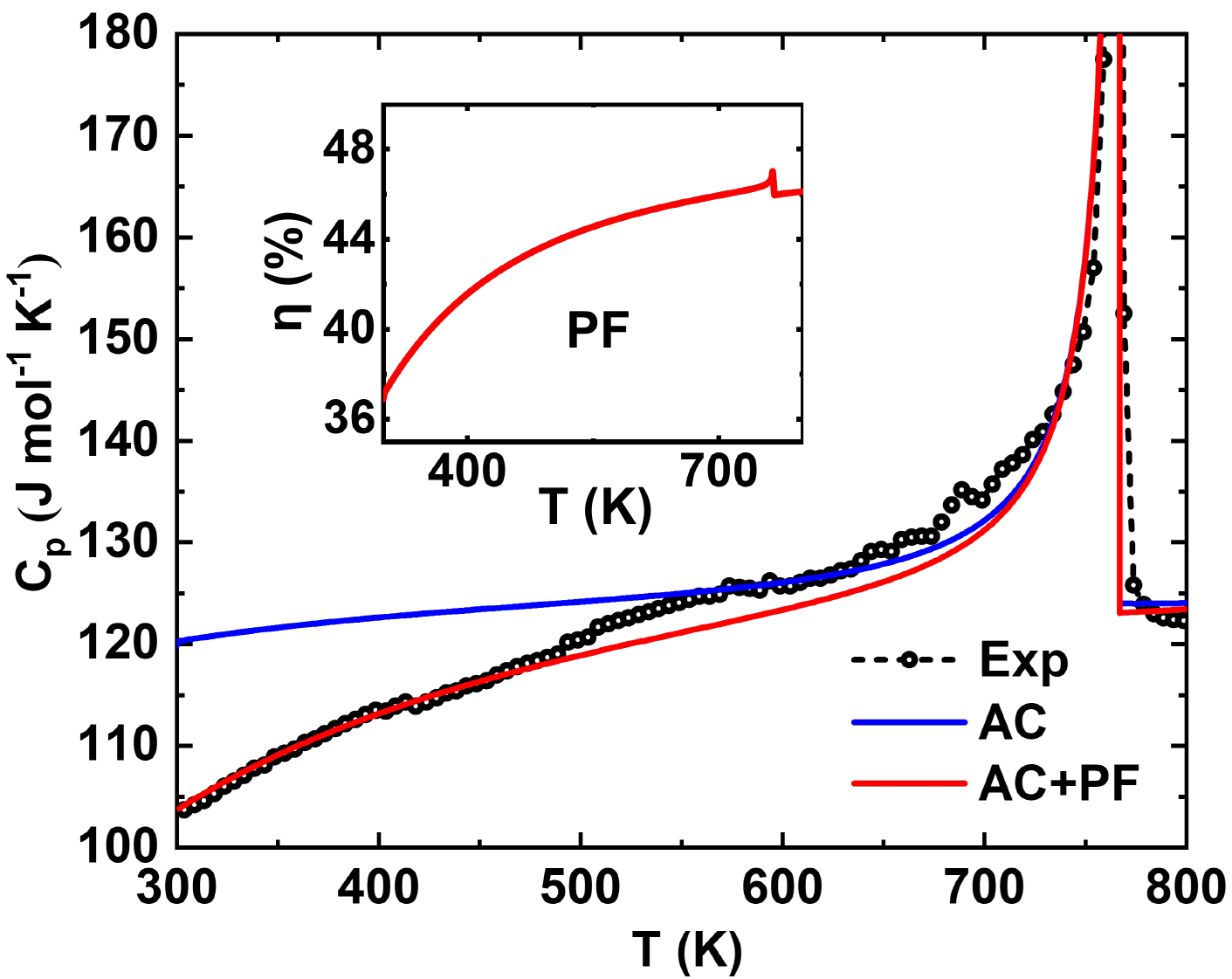}
    \caption{Molar specific heat $c_p$ of PbTiO$_3$ under ambient pressure, as compared to experimental data (black circles)~\cite{Yoshida2009}, is described by acoustic phonons (AC, blue curve) and by combined with  polarization fluctuations (AC+PF, red curve). 
    The inset indicates the ratio $\eta$ of ferron contribution to the total specific heat.
    }
    \label{fig:fig2}
\end{figure}

\begin{figure}[htbp]%[htbp]
\centering
\includegraphics[width=8.2cm]{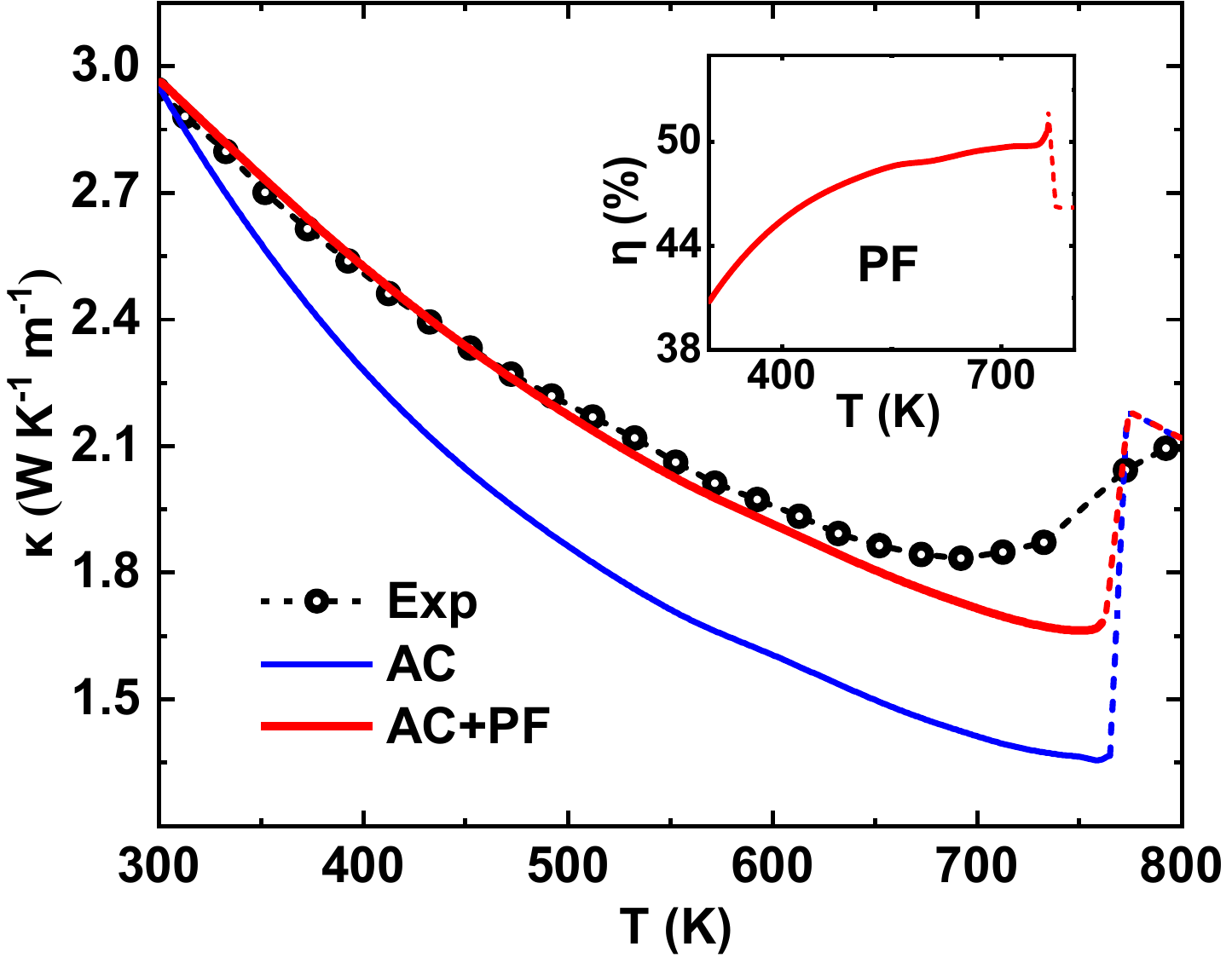}
    \caption{Thermal conductivity curves of PbTiO$_3$ from ceramic experiment (black dotted)~\cite{Tachibana2022}, described by acoustic phonon (blue curve) and after combined with  polarization fluctuations (red curve). 
    The red dash dot curve is calculated with adjusted $\tau$. 
    The inset indicates the ratio $\eta$ of ferron contribution in total thermal conductivity. 
    Dash-dot lines extended to the cubic PE phase are speculated with higher $\tau$.
    }
    \label{fig:fig3}
\end{figure}

%section_2-2
The predicted results from the full numerical calculation of the specific heat and thermal conductivity of ferroelectric PbTiO$_3$ are shown in Figs.~\ref{fig:fig2} and~\ref{fig:fig3}, respectively. 
Particularly, the molar specific heat at constant pressure in Fig.~\ref{fig:fig2} is related to the one at constant volume via the thermodynamic relation ${c_p-c_v=rN_A\alpha_{exp}^2 BVT}$~\cite{Chen2022}, where ${\alpha_{exp}}$ is the temperature-dependent volume expansion coefficient derived via Eq.~(\ref{eqn:volume}) from the experimental measurement~\cite{Haun1987}, $B$ is the bulk modulus~\cite{Jabarov2011} and $r$ is the number of atoms per unit cell.
Additionally, we have also consider the volume to be temperature-dependent, accounting for both thermal expansion and abnormal contraction due to polarization reduction (see Appendix~\ref{app:PbTiO$_3$}). 
Further details of the parameters used and their determination from independent measurements are provided in Table~\ref{tab:table1}.

%section_2-3
Fig.~\ref{fig:fig2} presents the temperature dependence of the specific heat at constant pressure $c_p$ in ferroelectric PbTiO$_3$, comparing the experimental measurement (black circles)~\cite{Yoshida2009} with theoretical predictions, including those from the classical acoustic phonon model (blue curve) and the extended model incorporating ferron contributions (red curve).
This comparison serves to highlight the limitation of the classical model and demonstrates how the new mechanism improves theoretical predictions.
The blue curve from the classical acoustic phonon model, i.e., Debye model, captures the general magnitude but significantly overestimate the thermal conductivity in the temperature range of $300$-$550\;\mathrm{K}$, failing to account for the experiment's obvious rising rate.
This discrepancy arises from the assumption of only acoustic phonon contributions in the classical model: 
in the absence of significant volume changes away from $T_c=765\;$K, it exhibits a nearly temperature-independent specific heat $c_p \! \approx \! c_v \! \approx \! 3N_Ak_B$ above its Debye temperature of $\sim 270\;$K, thus overlooking contributions from other  excitations.

%section_2-4
In contrast, the red curve from our extended model incorporating ferron contributions aligns closely with the experimentally measurements over a very wide temperature range, particularly capturing the rapid increase around $300\!-\!550\;\mathrm{K}$.
This improvement stems from the softening of ferron excitation energy at elevated temperatures, which enhances the low-energy contribution to heat capacity.
We now qualitatively analyse the temperature dependence of $c_v (\approx\!c_p)$ for ferrons around room temperature.
Assuming that the softening effect leads to a negligible ferron excitation gap and a constant group velocity, ferrons effectively behave like acoustic phonons, but with a larger Debye temperature 
($D \approx 850\;\mathrm{K}$, see examples in Appendix~\ref{app:wq}~\footnote{\label{mynote}Note that these discussions are based on unadjusted integration cut-off and are still valid in explaining the results with fitted parameters.}).
Under theses assumptions, the molar specific heat $c_v$ of ferron at room temperature can be approximated as
\begin{eqnarray}
    c_v
    &\propto&
    \frac{4 \pi^4 T^3}{15}
    -
    D^3 e^{-\frac{D}{T}} \!
    \left(
          12   \frac{T}{D}
        +  4   
        +      \frac{D}{T}
    \right)\!
    ,\label{eqn:serialcv}
\end{eqnarray}
where the first term corresponds to the low-energy contributions dominated by quasi-gapless states, being responsible for the characteristic $T^3$ scaling of acoustic-like modes;
the second term captures the suppression appearing in the high-temperature spectrum, governed by the exponential term $e^{-D/T}$.
As the temperature approaches the Debye temperature ($T\!\sim\!D$), the high-energy suppression term begins to play a significant role.
At intermediate temperatures e.g., near room temperature, the low-energy $T^3$-scaling still dominates the specific heat contribution, and there is still a rapid increasing rate of ferron specific heat (see more discussions in Appendix~\ref{app:pt}).
Thus, the inclusion of ferron contributions accounts for the experimentally observed rapid increase of $c_p$ between $300\!-\!550\;\mathrm{K}$, as shown in Fig.~\ref{fig:fig2}.
By properly incorporating the softening effects, this framework provides a more accurate description of the specific heat in ferroelectrics, highlighting the essential role of ferrons across a broad temperature range.

%section_2-5
The thermal conductivity of ferroelectric PbTiO$_3$ is shown in Fig.~\ref{fig:fig3}.
Although considering solely the contribution of acoustic phonons provides a qualitative description for the experimentally measured thermal conductivity, 
the theoretical prediction (red curve in Fig.~\ref{fig:fig3}) can quantitively capture the experimental data over a wide temperature range in the ferroelectric phase of PbTiO$_3$ only when both ferron and acoustic phonon contributions are included.
The inclusion of ferrons slows down the rate of decline in $\kappa$ with increasing temperature.
To gain a deep insight, we examine the relationship between specific heat and thermal conductibility.
Assuming the ferron quasi-particle gap $\omega_{q=0}$ becomes negligible due to softening, according to Eqs.~(\ref{eqn:cv})-(\ref{eqn:kappa}):
\begin{align}
    \kappa
    &=
    \frac{\tau v_g^2 (c_v - O)}{3 N_A V_0} 
    ,\label{eqn:kcv}
\end{align}
where $v_g^2=g/m_p$ and $O$ corresponds to the part of $c_v$ with $\partial \omega_q / \partial T$ in Eq.~(\ref{eqn:cv}) (see more details in Appendix~\ref{app:pt}). 
Neglecting the second term, Eq.~(\ref{eqn:kcv}) is simplified to $\kappa = \tau v_g^2 c_v / 3 N_A V_0$.
This expression mirrors the classical kinetic theory of gases~\cite{Kittel2005,Tritt2005}, reinforcing the interpretation that the thermal conductivity is jointly determined by the specific heat, group velocity, and relaxation time.
The ferron contribution to $\kappa$ stems from its significant impact on $c_v$.
The rapid increase in ferron specific heat at higher temperatures, driven by its softening gap and high effective Debye temperature, counterbalances the otherwise dominant $T^{-1}$ decay of $\tau$, thereby slowing the decline of $\kappa$.
This interplay between ferron dynamics and thermal transport not only explains the experimentally observed trends but also highlights the critical role of polarization fluctuations in determining the thermophysical properties of ferroelectrics.

%section_2-6
Notably, as shown in the inset of Fig.~\ref{fig:fig3}, the ferron contribution to $\kappa$ remains significant ($37\!\sim\!45 \%$), even though its relaxation time $\tau$ is an order of magnitude smaller than that of acoustic phonons.
This is primarily due to the softening of ferrons near the transition temperature, which reduces their quasi-particle gap $\omega_{q=0}$, allowing them to behave like acoustic phonons with negligible gaps.
With the sufficiently small gap, ferrons exhibit a higher group velocity $v_g$, approximately three times that of acoustic phonons. 
This elevated velocity amplifies their contribution to thermal conductivity, compensating for the shorter $\tau$, and making the overall thermal conductivity align well with the experimental observations.

%section_2-7
It is also noted that there is an obvious increase of $\kappa$ above $T_c$, as indicated by the dotted extensions in Fig.~\ref{fig:fig3}.
This increase is likely due to a significant reorganization of lattice acoustic phonon bands, which reflects the character of a first-order phase transition.
It is therefore necessary to assume a significantly reduced phonon-phonon interactions and hence increased $\tau$ after the phase transition above $T_c$, which is beyond the scope of this study.

\begin{table}[htbp]
\caption{\label{tab:table1}%
    Parameters for acoustic phonon (AC) and combined phonon and polarization fluctuation (AC+PF) models used in the calculations for Fig.~\ref{fig:fig2} and~\ref{fig:fig3}.
    Integration cut-off values are listed by their ratios to the Debye wavevector $q_D=({6\pi^2}\text{⁄}V_0)^{1\text{⁄}3}$. 
    In the combined AC+PF model, they are fitted to both the absolute value and changing rate of experimental $c_p$~\cite{Yoshida2009} and $\kappa$~\cite{Tachibana2022} around room temperature.
    Room temperature $\tau$ values in Fig.~\ref{fig:fig3} are close to literature values of 0.15 ps for PF~\cite{Tang2022,Sanjurjo1983} and 1.20 ps (extrapolated in doped PTO) for AC~\cite{Yurtseven2016}, respectively.}
\begin{ruledtabular}
\begin{tabular}{ccc}\label{TableI}
Models  &  $q~(\times q_D)$  &  $\tau_{RT}$~(ps)  \\ \hline
AC+PF   &  0.81, 1.14        &  1.00, 0.19 \\
AC      &  1                 &  1.09
\end{tabular}
\end{ruledtabular}
\end{table}

\begin{figure}[htbp]
\centering
\includegraphics[width=8.2cm]{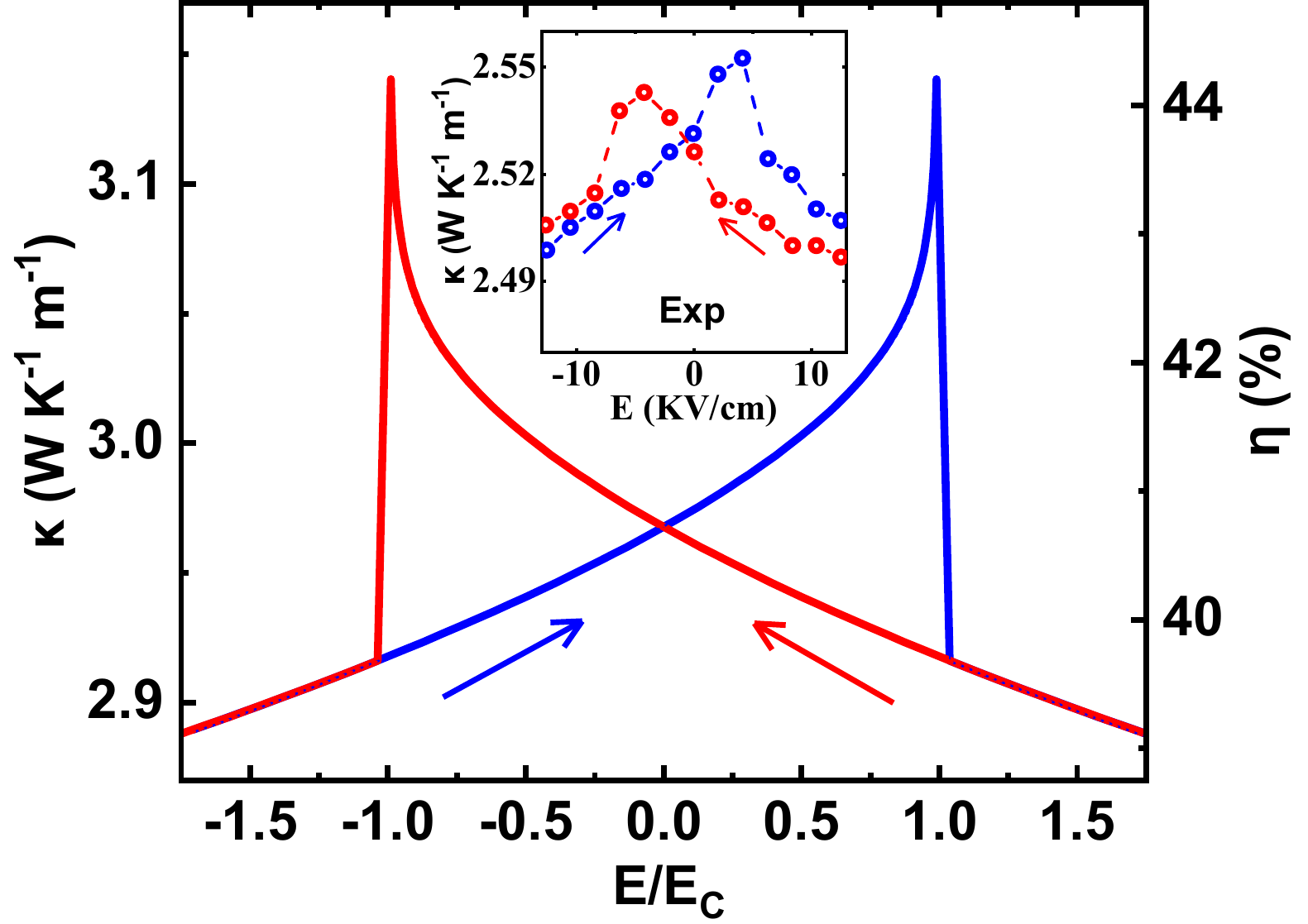}
    \caption{Thermal conductivity curves of PbTiO$_3$ under varying external electric field parallel with $P$ at room temperature, described by the combined model with both acoustic phonon and polarization fluctuation. 
    Inset shows an experimental $\kappa$ of lead zirconium titanate under electric field~\cite{Wooten2023}.
    }
    \label{fig:fig4}
\end{figure}

%section_2-8
Finally, we analyse the electric field dependence of the thermal conductivity at room temperature, which provides direct evidence for the significant ferron contribution to thermal transport. 
The external electric field $E$ directly affects the magnitude of the long-range ordered polarization $P_0$, the order parameter of ferroelectricity determined by the equilibrium condition
\begin{equation}
    \alpha P_0 + \beta P_0^3 + \lambda P_0^4 = E.
\end{equation}
This leads to the relation between $P_0$ and $\partial P_0^2\text{/}\partial T$, which reflects how the temperature dependence of polarization influences the pyroelectric coefficient, as
\begin{align}%\small
    \frac{\partial P_0^2}{\partial T} 
    = -2P_0^2 
    \left(
        \frac{\partial \alpha}{\partial T} 
    +   \frac{\partial \beta}{\partial T} P_0^2
    \right)
    \! \Big/ \!
    \left(
        m_p \omega_{q=0}^2
    \right)
    , \label{eqn:pps2pte}
\end{align}
where the temperature-sensitive quasi-particle energy gap $\omega_{q=0}^2=(\alpha + 3\beta P_0^2 + 5\lambda P_0^4)/m_p$ plays a central role.
Therefore, with the Landau free energy coefficients obtained through the self-consistent renormalization approach~\cite{yang2024microscopic}, we rigorously derive the electric field dependence of heat capacity and thermal conductivity in displacive ferroelectrics within quantum statistic theory.

The calculated $\kappa(E)$ curves at room temperature are presented in Fig.~\ref{fig:fig4}.
When the spontaneous polarization $P_0$ is reduced by an opposite external electric field~\cite{Landauer1956,Landauer1957,Bratkovsky2000}, the quasi-particle excitation gap $\omega_{q=0}$ decreases, mimicking the temperature softening effect approaching $T_c$.
This softening effect significantly enhances the population of low-energy ferrons, resulting in an increased ferron contribution to $\kappa$ (see detailed results in Appendix~\ref{app:pt}).
Notably, this effect is not confined to the vicinity of $T_c$: the tunable energy gap via the electric field extends the influence of ferrons over a wider temperature range.
The changes in the dispersion relation further amplifies this effect:
as $\omega_{q=0}$ decreases, the entire branch of the dispersion relation $\omega_q$ shifts downward following Eq.~(\ref{eqn:wq}), leading to an increased density of states for low-energy ferrons.
Such behaviours enhance the specific heat contribution and, consequently, the thermal conductivity, aligning with experimentally observed trends where external electric fields modulate thermal transport~\cite{Wooten2023}.
Thus, the field dependence of $\kappa$ underscores the critical role of ferrons in ferroelectric thermal transport.
The ability of ferrons to exhibit field-tunable softening offers a promising pathway to understanding and leveraging the interplay between electric fields and thermal transport properties in ferroelectrics.
These findings open avenues for designing innovative energy and thermal management technologies using ferroelectric materials.

%Section_3-1
\begin{figure}[htbp]
\centering
\includegraphics[width=8.2cm]{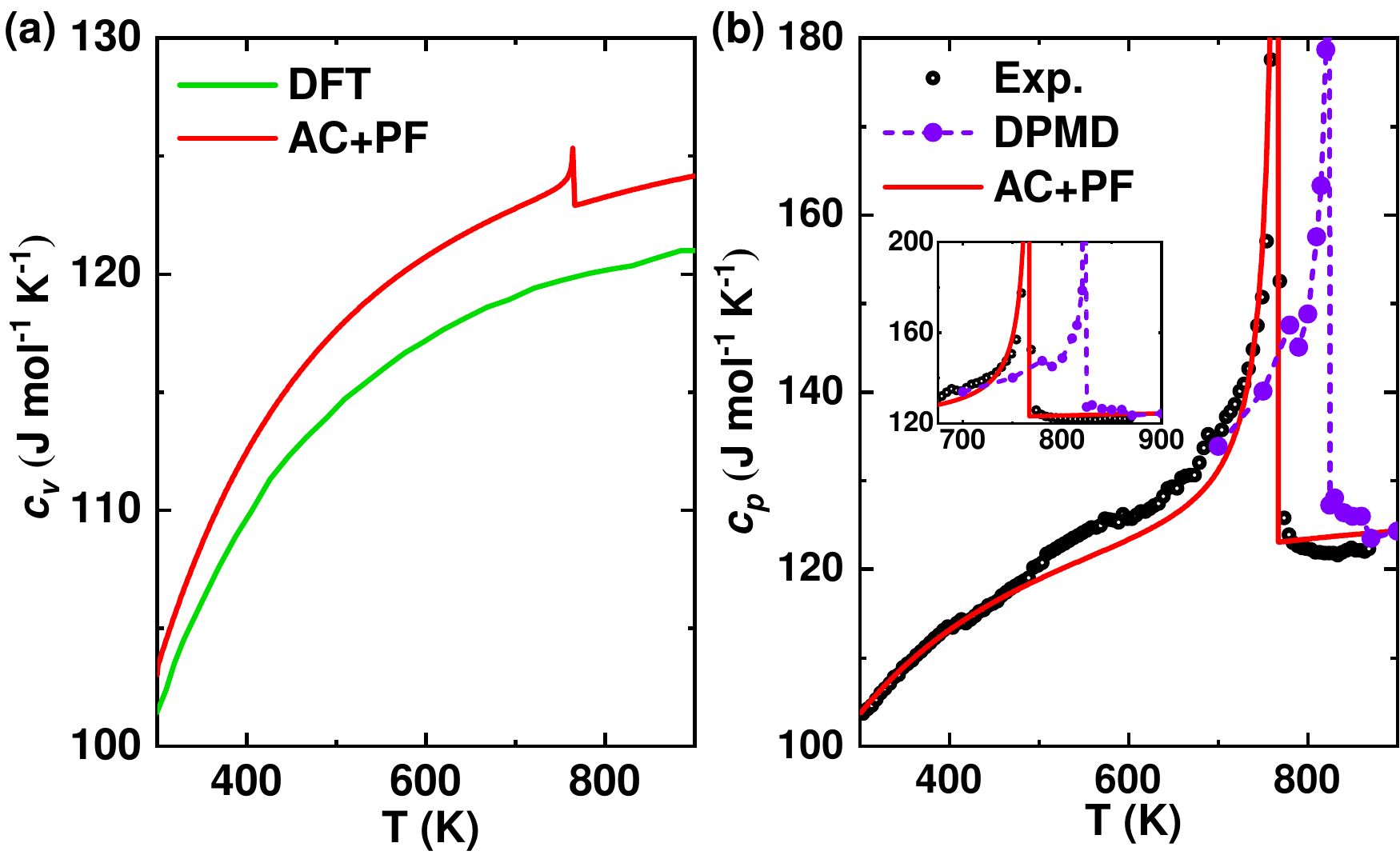}
    \caption{(a) Temperature-dependent molar specific heat at constant volume, $c_v$, of PbTiO$_3$: comparison between our statistical model with both acoustic phonons and polarization fluctuations (AC+PF, red curve, with $T_c\approx765\;$K) and DFT simulation result of Z. Wang et al. (DFT, green curve)~\cite{Wang2021}. 
    (b) Temperature-dependent molar specific heat at constant pressure, $c_p$, of PbTiO$_3$: comparison among experimental data from T. Yoshida et al. (Exp., black circles, with $T_c\approx765\;$K)~\cite{Yoshida1960}, deep potential molecular dynamics (DPMD, purple dots, with $T_c\approx821\;$K) trained on DFT-based data~\cite{Xie2025}, and our statistical model (AC+PF, red curve, with $T_c\approx765\;$K). The inset shows a magnified view.
    }
    \label{fig:fig5}
\end{figure}

\textit{Comparison with other Methods.---}In Fig.~\ref{fig:fig5}(a), we compare the molar specific heat at constant volume $c_v$ of PbTiO$_3$. 
Our model with both acoustic phonons and polarization fluctuations contributions (AC+PF) successfully reproduces the discontinuous jump at the Curie temperature, reflecting the first-order nature of the ferroelectric phase transition. 
In contrast, existing DFT-based methods, such as the calculation by Wang et al.~\cite{Wang2021}, underestimate the heat capacity and also fail to capture this discontinuity due to the absence of phase transition.

%Section_3-2
In Fig.~\ref{fig:fig5}(b), we present the comparison of molar specific heat at constant pressure $c_p$. 
Our model again shows excellent agreement with experimental data from Yoshida et al.~\cite{Yoshida2009}. 
Meanwhile, the state-of-the-art deep potential molecular dynamics (DPMD) simulation~\cite{Xie2025} still overestimates the transition temperature ($T_c \approx$ 821 K).
Ab initio molecular dynamics (AIMD) simulations~\cite{Cockayne2018}, as an alternative basis of learning finite-temperature lattice dynamics, often require empirical correction of lattice parameters to match experiments in PbTiO$_3$, and various of exchange-correlation functionals are still not ideal in reproducing c/a ratio. 
AIMD often underestimates $T_c$ by more than 100 K~\cite{Nishimatsu2012,Wang2023}.
These comparisons emphasize the key strength of our approach: 
it incorporates low-energy collective excitations, ferrons, with their temperature-dependent softening, allowing accurate and physically transparent modeling of thermodynamic properties near the ferroelectric transition, which is difficult to achieve with current first-principles-based simulations.

\begin{figure}[htbp]
\centering
\includegraphics[width=8.2cm]{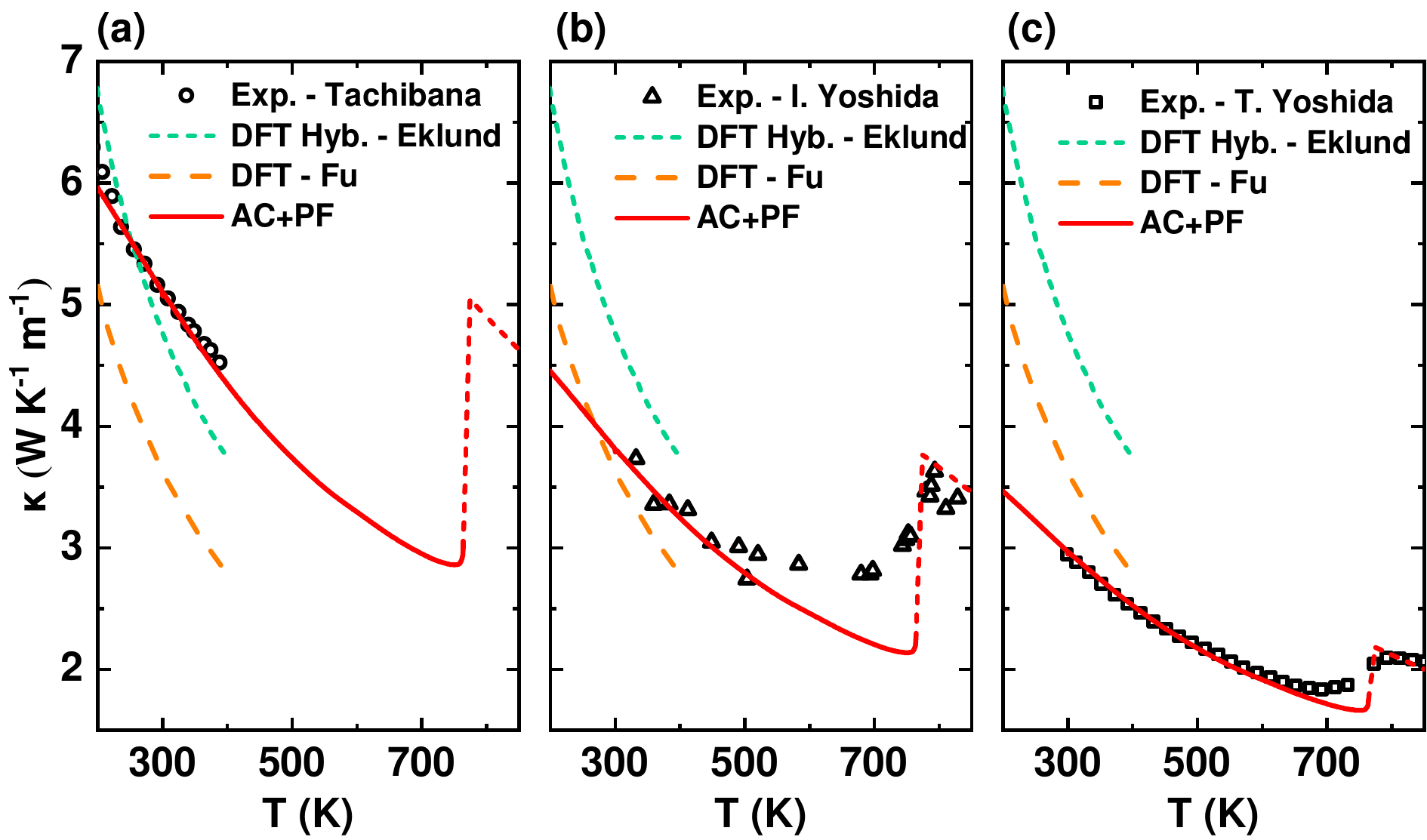}
    \caption{Comparison of average temperature-dependent thermal conductivity $\kappa$ of PbTiO$_3$ from experiments~\cite{Tachibana2008}, I. Yoshida et al.~\cite{Yoshida1960}, and T. Yoshida et al.~\cite{Yoshida2009} and calculations: DFT-LDA with 0 K phonon dispersion by Y. Fu et al.~\cite{Fu2018} (orange dashes); DFT hybrid (PBEsol0) with self-consistent phonon (SCPH) theory by K. Eklund et al.~\cite{Eklund2024} (green short dashes); statistical models with both acoustic phonons and polarization fluctuations (AC+PF, red lines), fitted only to the $\tau$ at 300 K of each experimental dataset. Curves in the paraelectric phase are speculated with $\tau$ from the T. Yoshida et al. data \cite{Yoshida2009}.
    }
    \label{fig:fig6}
\end{figure}

%Section_3-3
As shown in Fig.~\ref{fig:fig6}, our model combining both acoustic phonons and polarization fluctuations (ferrons) successfully reproduces the temperature-dependence trends of thermal conductivity across multiple experimental datasets, with all input parameters consistently taken from experiments or our self-consistent renormalization theory~\cite{yang2024microscopic}, ensuring physical reliability especially near the ferroelectric transition.
Although these experiments differ in absolute $\tau$ values, they share nearly identical temperature slopes, indicating that the variation mainly stems from differences in the overall relaxation time scale.
We adopt a temperature-dependent form $\tau(T) \propto 1/T$, motivated by acoustic phonon populations. 
While this approximation may break down very close to $T_c$ due to strong anharmonicity, it suffices to capture global trends without modeling detailed scattering:
by fitting the relaxation time $\tau_{RT}$ at a single reference temperature (300 K) and adopting a fixed ratio of acoustic phonon and ferron contributions (as listed in Table~\ref{tab:table1}), our model can be rescaled to match each dataset using only one parameter. 
Once ferron softening is correctly captured, variations in magnitude are readily adjusted via a simple $\tau$ rescaling (1.72~\cite{Tachibana2008} and 1.29~\cite{Yoshida1960} relative to the $\tau$ used for~\cite{Yoshida2009}).
This highlights both the robustness and efficiency (less than 1 minute per full temperature curve on 24 cores) of our approach in describing thermal transport in displacive ferroelectrics.

%Section_3-4
We note that first-principles methods using static 0 K phonons (e.g., Fu et al.~\cite{Fu2018}) lack soft-mode renormalization, while self-consistent phonon (SCPH) based approach with hybrid functionals (e.g., Eklund et al.~\cite{Eklund2024}) still show notable deviation from experiment.
These comparisons suggest that it is the lack of proper description of soft-mode softening and polarization fluctuations, not the $\tau \propto 1/T$ approximation or incomplete phonon dispersion, that fundamentally limits the predictive accuracy.

%Summary
\textit{Summary and Discussion.---}We present a theoretical framework to analyse the role of ferrons—elementary excitation of polarization—in the thermal properties of displacive ferroelectrics.
By employing a self-consistent renormalization approach, we incorporated the temperature-dependent softening of ferron excitation energy, quantitatively capturing key experimental observations of specific heat and thermal conductivity in PbTiO$_3$.
Our results demonstrate the pivotal role of ferrons in determining the thermal transport near the phase transition, where their dispersion exhibits pronounced sensitivity to both temperature and external electric fields.
This study bridges the gap between classical phonon models and experimentally observed thermal anomalies in ferroelectrics.
Notably, we demonstrated how the softening feature of ferrons can account for their significant contributions to heat capacity and thermal conductivity, especially in scenarios involving strong external field effects. 
Our study provides a quantitatively accurate and computational efficient description of finite-temperature thermodynamics based on ground-state information, bridging the gap between classical phonon models and experimentally observed thermal anomalies in ferroelectrics.
The proposed framework here is extendable to a wide range of ferroelectric materials as ferrons universally serve as low-energy excitations near ferroelectric phase transitions, where their softening leads to significant thermal contributions similar to acoustic phonons. This provides, insights into polarization-driven thermodynamics and guides the development of energy and thermal management systems.

Regarding the limitations of our model, there are some aspects worth to know about: 
First, the present study focusing on the essential low-lying modes does not perform full phonon dispersion calculations in the entire Brillouin zone~\cite{TDEP1,TDEP2,TDEP3,TDEP4,TDEP5,TDEP6,TDEP7}, as in the usual thermal-transport calculation of phonons. This is because fundamental physics in solid-state theory and quantum statistical mechanics dictate that in bosonic systems, low-energy excitations dominate thermodynamic properties and transport behaviours, in contrast to fermionic systems, where contributions from the entire Brillouin zone can be significant.
This approximation starts to lose accuracy at very high temperatures when other optical phonon modes are sufficiently excited.

Second, it should be emphasized that our study focuses on the homogeneous bulk ferroelectric phase transition, where the long-wavelength fluctuations should dominate over a broad temperature range above zero, and hence high-order terms such as flexoelectric effects are neglected. 
However, as pointed out recently~\cite{Dieguez2022}, a consistent thermodynamic formulation may require the inclusion of both flexoelectric effect and polarization gradient energy in highly inhomogeneous systems.
A key objective of the present work is to determine the amount and significance of ferron contributions to the overall thermal properties.
Constructing an inhomogeneous model that integrates the self-consistent phase-transition theory would provide a more comprehensive theoretical framework, but this is beyond the scope of the present study.
We leave this for future research, particularly in the context of domain walls, which are known to influence thermal transport via phonon scattering~\cite{Mante1971}.

Third, in comparison with conventional phonon-based approaches, our method offers a unique perspective by explicitly incorporating ferron softening $\partial \omega/\partial T$ in the drive term of the Boltzmann transport equation, enhancing predictions of thermal conductivity near phase transitions. 
Based on the relaxation time approximation, our approach does not explicitly include phonon-phonon interactions~\cite{Fu2018,Cazorla2024}, and this will lead to errors close to $T_c$ where profound anharmonicity emerges. 
Future first-principles calculations incorporating a fully microscopic and complete bosonic scattering treatment can fill the gap here by providing the specific relaxation time value.

%Acknowledgeent
\textit{Acknowledgement.---}This work was supported as part of the Computational Materials Sciences Program funded by the U.S. Department of Energy, Office of Science, Basic Energy Sciences, under Award No. DE-SC0020145. 
G.D.Z., F.Y., and L.Q.C. also appreciate the generous support from the Donald W. Hamer Foundation through a Hamer Professorship at Penn State.

%--------------------------------------------------------------------------
%                                 Appendix
%--------------------------------------------------------------------------
\appendix
\setcounter{figure}{0}
\renewcommand{\thefigure}{A\arabic{figure}}

\section{\label{app:wq}
    \texorpdfstring{Dispersion relation of polarization fluctuation in $\text{PbTiO}_3$}{Section-1:~Dispersion relation of polarization fluctuation in PbTiO3}}

\begin{figure}[htbp]%h is too strict
\centering
\includegraphics[width=8.2cm]{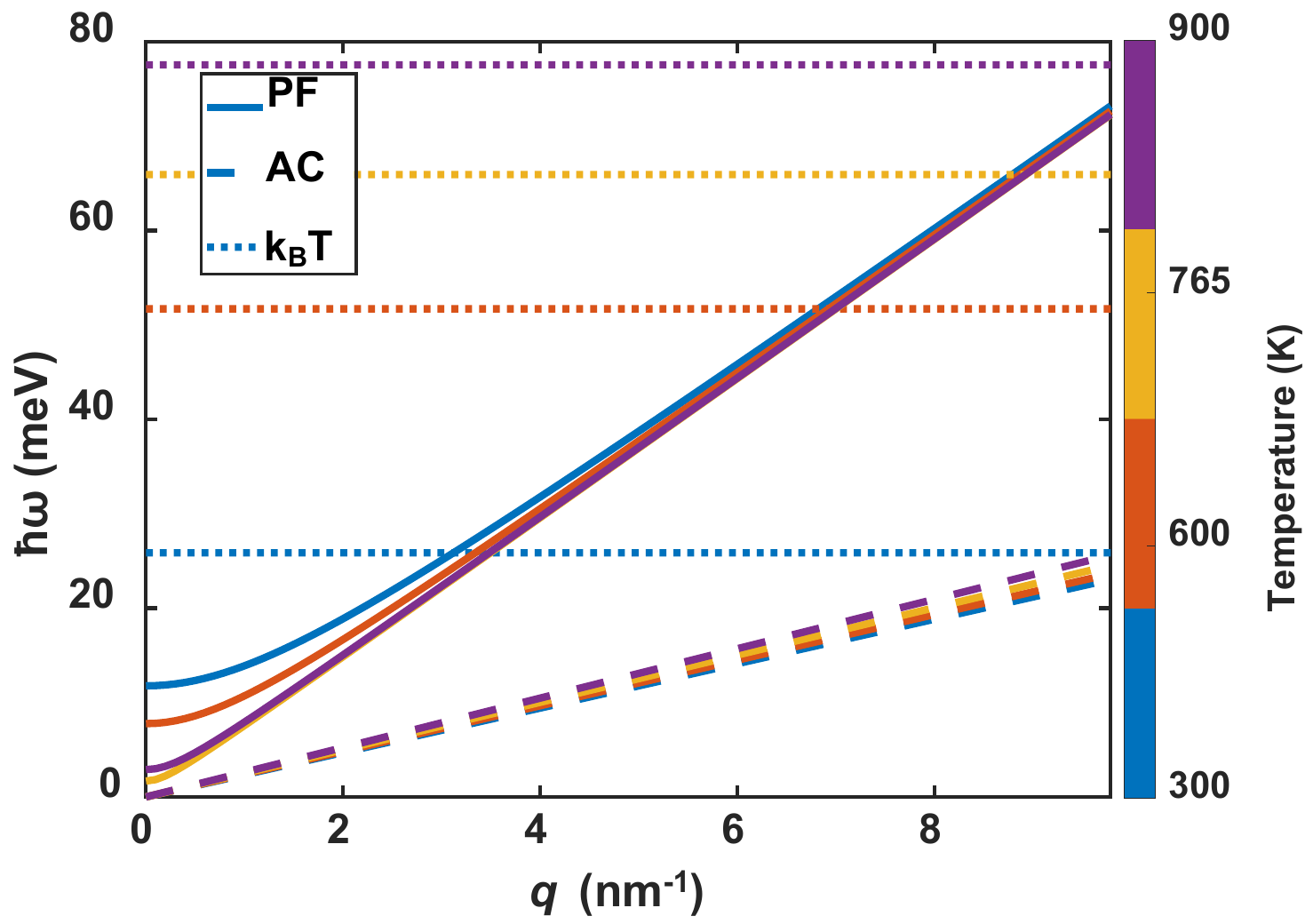}
    \caption{Dispersion relation of polarization fluctuation (PF) and acoustic phonon (AC) at some chosen temperatures across $T_c$ in PbTiO$_3$.
    }
    \label{fig:figs1}
\end{figure}

As shown in Fig.~\ref{fig:figs1}, the dispersion relation of polarization fluctuation becomes approximately linear at ${T_c=765\;\mathrm{K}}$, together with a minimum energy gap $\omega_{q=0}$ at $\lambda$-point. 
For temperatures away from $T_c$, the gap and group velocity gradually deviate from their extreme values.
Note the predicted $\omega_{q=0}$ by self-consistent renormalization group theory at room temperature is consistent with experimental measurements~\cite{Hlinka2006}. 

\begin{figure}[b]
%\centering
\raggedright
\includegraphics[width=8.0cm]{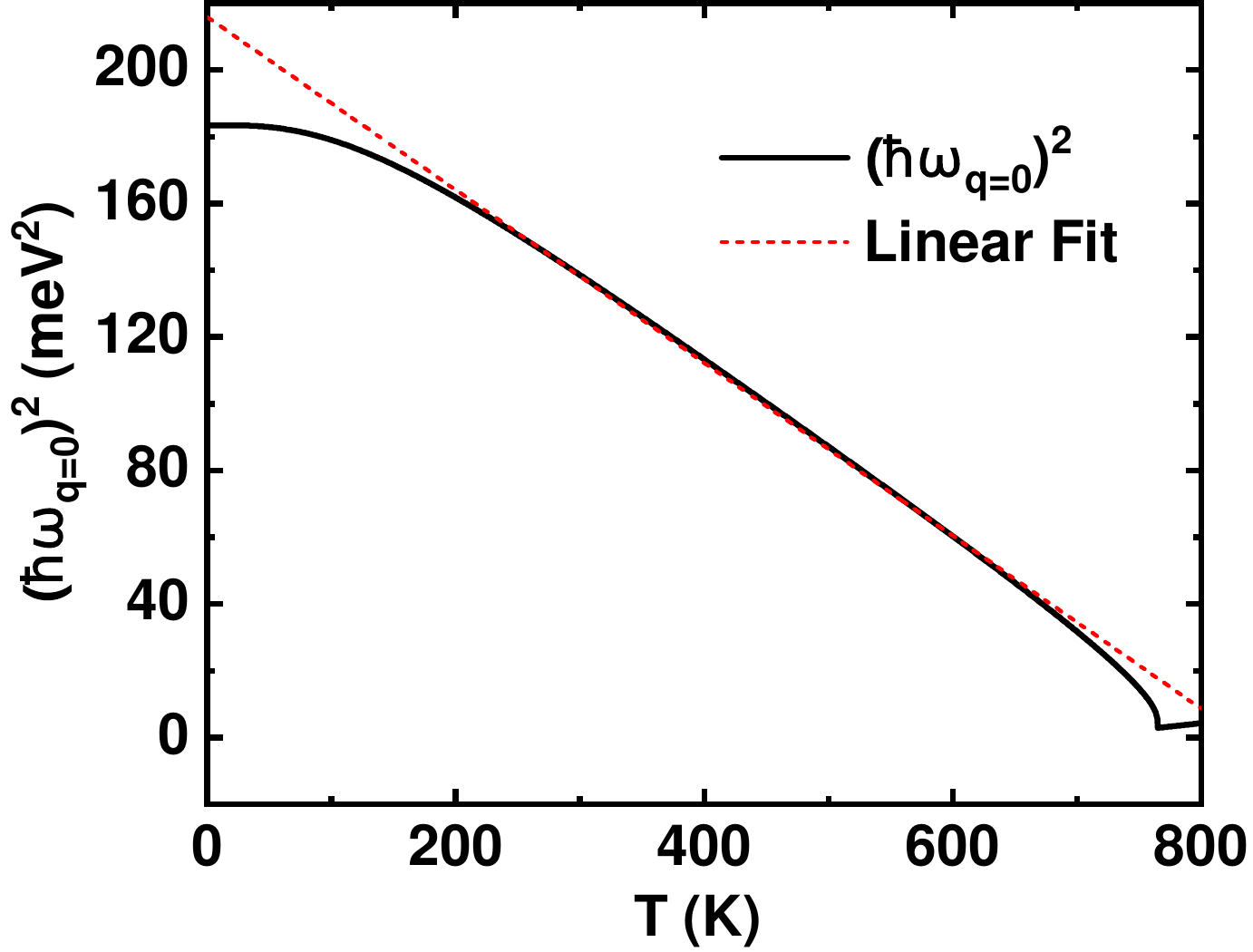}
    \caption{The squared quasi-particle gap $\omega_{q=0}$ of ferron is displayed from zero temperature to $T_c$ of PbTiO$_3$, with renormalized parameters.
    }
    \label{fig:figs2}
\end{figure}

As shown in Fig.~\ref{fig:figs2}, the squared quasi-particle gap $(\hbar \omega_{q=0})^2=(\alpha+3\beta P_0^2 +5\lambda P_0^4)/m_p$ of ferrons basically follows a linear temperature dependence around/above room temperature, reaching a finite minimum at $T_c$ and converges at zero temperature.
The quasi-linear regime is fitted by the function, $(\hbar \omega_{q=0})^2=a-bT$, where $a=216\;\mathrm{(meV)^2/K}$ and $b=0.259 \;\mathrm{(meV)^2/K}$.

\begin{figure}[htbp]%h is too strict 
\centering
\includegraphics[width=8.0cm]{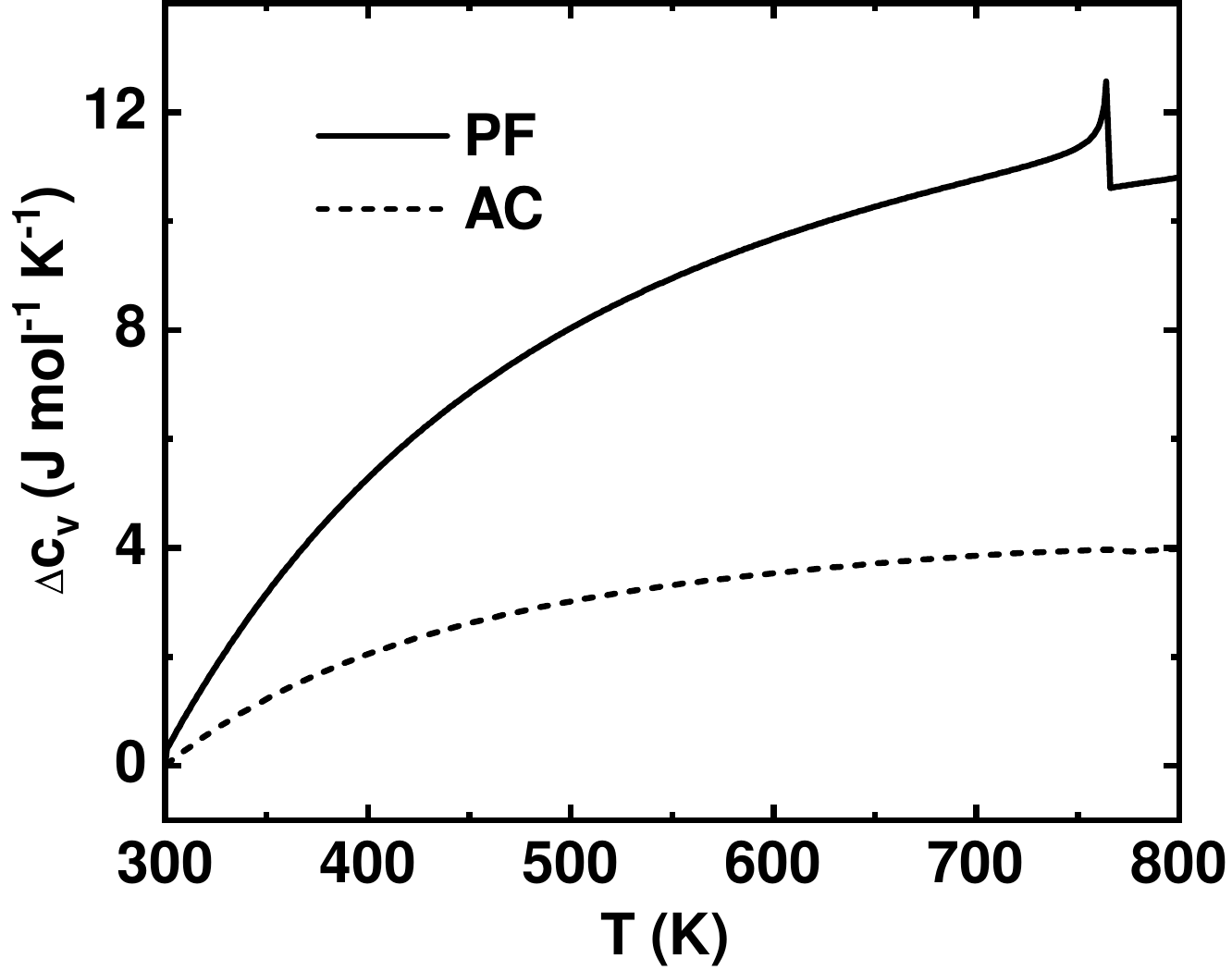}
    \caption{The change of specific heat, $\Delta c_v$, from room temperature to $T_c$ of polarization fluctuation (PF) and acoustic phonon (AC) in PbTiO$_3$, both integrated to the Debye wavevector $q_D$.
    }
    \label{fig:figs3}
\end{figure}

As compared in Fig.~\ref{fig:figs3}, the change of molar specific heat $c_v$ in the range of $T_{RT} \sim T_c$ is mainly contributed by the polarization fluctuation.
In other words, the $c_v$ of polarization fluctuation quasi-particle varies quicker in $T_{RT} \sim T_c$ than acoustic phonons.

\section{\label{app:pt}Temperature-dependent specific heat and thermal conductibility}

We first briefly review the textbook derivations of the conventional Debye’s phonon model~\cite{Kittel2005,Callaway2013}, which is appropriate to describe the thermal heat capacity and thermal transport dominated by the acoustic mode, especially at low temperatures and long wavelength. 
The dispersion of acoustic phonons is  ideally linear as in a continuum elastic medium:
\begin{align}%{\small}
    \omega = v_g q,\label{eqn:debye}
\end{align}
with $v_g$ being the group velocity. The corresponding total energy of lattice vibrations is
\begin{align}
    u
     = \int_0^{\omega_D} d\omega\; D(\omega) \hbar \omega \bar{n}_q 
    =\frac{V_0 \hbar v_g}{2\pi^2} 
      \int_0^{q_D} dq\; q^3 \bar{n}_q.
      \label{eqn:debyeU}
\end{align}
Considering three branches of acoustic phonons with average sound velocity $v_g$, the molar specific heat is 
\begin{equation}
    c_v 
     = 3 N_A \left(\frac{d u}{d T}\right)_V =\frac{3 N_A V_0 \hbar v_g}{2\pi^2 k_B T^2} 
      \int_0^{q_D} dq\; q^4 \bar{n}_q (\bar{n}_q + 1).
      \label{eqn:debyecv}
\end{equation}
where the temperature derivative of the Bose-Einstein distribution is simply
\begin{align}
    \frac{\partial \bar{n}_q}{\partial T} = \frac{\hbar \omega}{k_B T^2} \bar{n}_q (\bar{n}_q + 1).
\end{align}
The integration radius upper limit in the reciprocal space is determined as Debye wavevector $q_D=({6\pi^2}\text{⁄}V_0)^{1\text{⁄}3}$, according to the total number of degrees of freedom in N unit cells per branch of phonon $\int_0^{q_D} dq\; v_g D(\omega) =N$. 
Thermal conductivity can be obtained using the kinetic theory of gases as:
\begin{align}
    \kappa = \frac{\hbar^2 v_g^4 \tau}{2\pi^2 k_B T^2} \int_0^{q_D} dq\; q^4 \bar{n}_q (\bar{n}_q + 1).
\end{align}

Now, we turn to the thermal properties of polarization fluctuations. 
According to Eq.~(\ref{eqn:wq}) in the main text, its density of states is
\begin{align}
    D(\omega_q)=\frac{dN}{d\omega_q}
               =\frac{V_0 m_p}{2 \pi^2 g} q \omega_q,
\end{align}
and therefore, total energy of polarization fluctuations is 
\begin{align}
    u
    = \int_0^{\omega_D} d\omega_q\; D(\omega_q) 
       \hbar \omega_q \bar{n}_q = \frac{V_0 \hbar}{2\pi^2} \int_0^{q_D} dq\; q^2 \omega_q \bar{n}_q,
    \label{eqn:totalU}
\end{align}
which has no fundamental difference from Eq.~(\ref{eqn:debyeU}) and can be used to calculate the specific heat in Eq.~(\ref{eqn:cv}) in the main text. 
Specifically, the molar total energy in a crystal with $r$ atoms in the primitive cell is 
\begin{eqnarray}
    u_m
    &=& 
    \frac{r N_A V_0 \hbar}{2\pi^2} \frac{m_p}{g}
    \int_{\omega_{q=0}}^{\omega_D} d\omega_q\; q \omega_q^2 \bar{n}_q =
    \frac{r N_A V_0 \hbar}{2\pi^2}
    \left( \frac{m_p}{g} \right)^{3/2}
    \nonumber\\
    &&\mbox{}\times\left( \frac{k_B T}{\hbar} \right)^4 
    \int_{x_0}^{x_D} dx\; x^2 \sqrt{x^2-x_0^2} \; \bar{n}_q
    ,\label{eqn:totalu}
\end{eqnarray}
where $x=\hbar\omega_q\text{/}(k_B T)$, and $x_0=\hbar\omega_{q=0}\text{/}(k_B T)$ with quasi-particle gap $\hbar\omega_{q=0} = \hbar[(\alpha + 3\beta P_0^2 + 5\lambda P_0^4)\text{/}m_p]^{1\text{/}2}$. 
$\mathrm{V_0}$ denotes the unit cell volume, $\hbar$ the reduce Plank constant, $\mathrm{k_B}$ the Boltzmann constant, $\mathrm{N_A}$ the Avogadro constant, and $\bar{n}_q=\left[\exp(\hbar\omega_q/(k_B T))-1\right]^{-1}$ the Bose-Einstein distribution at thermal equilibrium. 
The $x_0 \neq 0\;$ except at $T_c$ in a second-order transition. 
Its existence rectifies the thermal behaviour of polarization fluctuation and deviates it from acoustic phonon's in Eq.~(\ref{eqn:debyeU}).

Substituting the term ${\partial \bar{n}_q}\text{⁄}\partial T$ into Eq.~(\ref{eqn:totalU}), the expression of $c_v$ and $\kappa$ in molar writes
\begin{widetext}
\begin{align}
    c_v 
    &=  \frac{r N_A V_0\hbar}{2\pi^2}
        \int_0^{q_D} dq\; q^2
        \left[ \frac{\partial \omega_q}{\partial T} \bar{n}_q +
            \frac{k_B}{\hbar} \frac{\hbar \omega_q}{k_B T} 
            \left(\frac{\hbar \omega_q}{k_B T} - 
                 \frac{\hbar}{k_B}\frac{\partial \omega_q}{\partial T}
                 \bar{n}_q (\bar{n}_q + 1)
            \right)
        \right],  \\
    \kappa 
    &=  \frac{r \tau g k_B}{6\pi^2 m_p} 
        \int_0^{q_D} dq\; \frac{q^4}{\omega_q^2}
        \left[ \frac{\hbar \omega_q}{k_B T} 
            \left(\frac{\hbar \omega_q}{k_B T} -\frac{\hbar}{k_B} 
            \frac{\partial \omega_q}{\partial T}
            \right) \bar{n}_q (\bar{n}_q + 1)
        \right].
\end{align}
\end{widetext}

A full expression of Eq.~(\ref{eqn:kcv}), where we neglect the quasi-particle gap $\omega_{q=0}$ and assume a constant squared group velocity $v_g^2 = g/m_p$, is
\begin{align}
    \kappa   
    &=
    \frac{\tau v_g^2}{3 N_A V_0}
    \left(
    c_v - 
    \frac{N_A V_0 \hbar}{2\pi^2}
    \int_0^{q_D} dq \; q^2 \frac{\partial \omega_q}{\partial T} \bar{n}_q
    \right)
    .%\label{eqn:kcv}
\end{align}
For ferrons with zero gap and constant group velocity (when it is like a branch of acoustic phonon), we formulate their molar total energy in the explicit form of serial number with their Debye temperature as the only parameter, as
\begin{align}
    u_m
    &=  \frac{N_A V_0 \hbar}{2\pi^2 v_g^3} 
        \left( \frac{k_B T}{\hbar} \right)^4
        \int_0^{x_D} dx\; x^3 \bar{n}_q
      ,\label{eqn:linearu}
\end{align}
where $x_D=D/T$ with ferron's Debye temperature $D\approx850\;\mathrm{K}$.
Replacing $\bar{n}_q=\sum_{s=1}^{\infty} e^{-sx}$, and with the relation in integration table
\begin{align}
    \int x^n e^{ax} dx 
    = \frac{1}{a} x^n e^{ax} - \frac{n}{a} \int x^{n-1} e^{ax} dx,
\end{align}
we have
\begin{eqnarray}
    u_m 
    &\propto&
        T^4  \sum_{s=1}^{\infty}
        \int_0^{x_D} dx\; x^3 e^{-sx}=\frac{\pi^4 T^4}{15}
        \!-\!\!\sum_{s=1}^{\infty} 
        e^{-sx_D}      \nonumber\\
    && 
                \times\Big(
         \frac{D^3T}{s}\!+\!\frac{3D^2T^2}{s^2}       
        \!+\!\frac{6DT^3}{s^3}
        \!+\!\frac{6T^4}{s^4}
        \Big), 
      %\label{eqn:linearu}
\end{eqnarray}
and correspondingly 
\begin{eqnarray}
    c_v 
    &=& \frac{d u_m}{d T} \nonumber\\
    &=&  
        \frac{N_A V_0 \hbar}{2\pi^2 v_g^3} 
        \left( \frac{k_B}{\hbar} \right)^4 
        \Big[
        \frac{4 \pi^4 T^3}{15}
        -
            \sum_{s=1}^{\infty} 
        e^{-\frac{sD}{T}}
        \Big(\!
          \frac{24}{s^4}         T^3   \nonumber\\
    & &
        + \frac{24}{s^3}   D    T^2
        + \frac{12}{s^2}   D^2  T
        + \frac{ 4}{s  }   D^3  
        +                  D^4  T^{-1}
        \Big) \Big]                    \nonumber\\
    &\propto&
    \frac{4 \pi^4 T^3}{15}
    -
    \sum_{s=1}^{\infty} 
        e^{-\frac{sD}{T}}
        \Big(\!
          \frac{24}{s^4}         T^3 + \frac{24}{s^3}   D    T^2
        + \frac{12}{s^2}   D^2  T  \nonumber\\
    & &        
        + \frac{ 4}{s  }   D^3  
        +                  D^4  T^{-1}
        \Big)     \propto   \frac{4 \pi^4 T^3}{15}
    -
    D^3
    \sum_{s=1}^{\infty} 
        e^{-\frac{sD}{T}}
        \Big(\!
          \frac{24}{s^4}   \frac{T^3}{D^3}                   \nonumber\\
    & &
        + \frac{24}{s^3}   \frac{T^2}{D^2}
        + \frac{12}{s^2}   \frac{T  }{D  }
        + \frac{ 4}{s  }     
        +                  \frac{D  }{T  }
        \Big)
      .%\label{eqn:linearcv1}
\end{eqnarray}
When $T$ is around room temperature with large $D/T$, we can neglect the terms with $s\gg1$ and the first two terms in the brackets, thereby give
\begin{eqnarray}
    c_v
    &\propto&
    \frac{4 \pi^4 T^3}{15}
    -
    D^3
    e^{-\frac{D}{T}} \!
    \left(
              12   \frac{T}{D}
        \!+\!  4   
        \!+\!      \frac{D}{T}
    \right)\!
    .%\label{eqn:linearcv1}
\end{eqnarray}
Here one can estimate the temperature scaling of $c_v (\approx\! c_p)$ at room temperature by taking $D/T\!\sim\!3$, obtaining rapid increasing rate as expected.

\begin{figure}[htbp]%h is too strict 
\centering
\includegraphics[width=8.2cm]{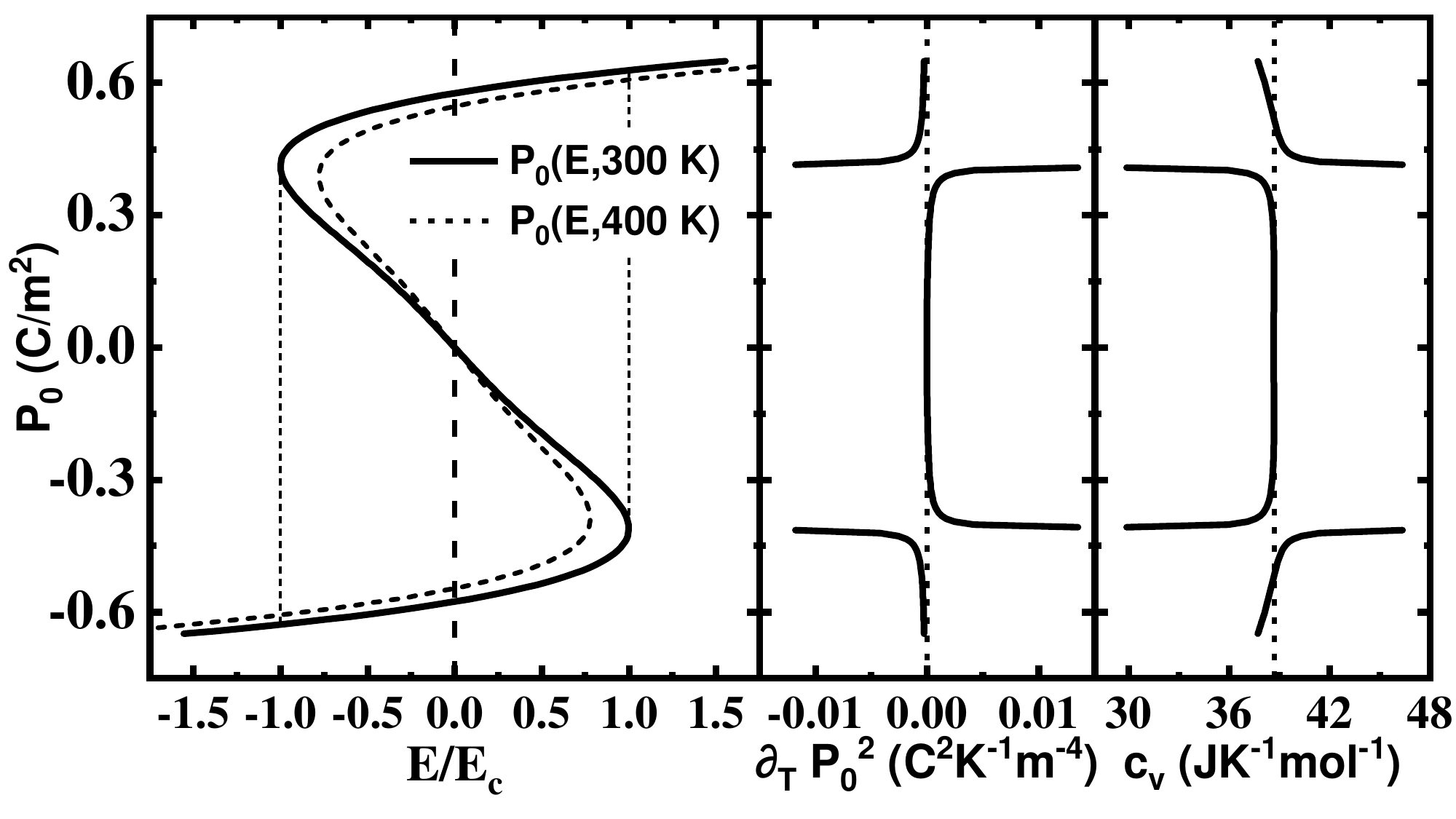}
    \caption{From left to the right are respectively: 
    E-P relation under equilibrium calculated according to Eq.~(\ref{eqn:ep}) at 300$\;$K (and 400$\;$K),
    $\partial P_0^2/\partial T$, and $c_v$ of ferron with integration cut-off of $1.14 \; q_D$
    .}
    \label{fig:figs4}
\end{figure}

For the temperature derivative of spontaneous polarization at equilibrium, the first step is to obtain the expression of 
\begin{align}%\small
    \frac{\partial F_P}{\partial P}
        = \alpha P + \beta P^3 + \lambda P^4 - E
        = 0
        , \label{eqn:ep}
\end{align}
which is the relation between an external electric field and spontaneous polarization $P_0$.
When $E=0$, we have the explicit expression in Eq.~(\ref{eqn:P0}).
When $E\neq0$, numerical calculations are required, and we need the relation between $P_0$ and $\partial P_0^2\text{/}\partial T$, where the former is directly known at each temperature and each electric field based on Eq.~(\ref{eqn:ep}).
We differentiate both sides of Eq.~(\ref{eqn:ep}) with respect to T, and we have
\begin{align}%\small
    \partial_T (\alpha P) + \partial_T(\beta P^3) + \partial_T(\lambda P^4)= 0
    ,% \label{eqn:peppt}
\end{align}
which can be expanded and rearranged as
\begin{align}%\small
    \frac{\partial P_0}{\partial T}
    = -2P_0 
      \frac{\frac{\partial \alpha}{\partial T} 
    + \frac{\partial \beta}{\partial T} P_0^2}
    {\alpha + 3\beta P_0^2 + 5\lambda P_0^4}
    ,% \label{eqn:peppt}
\end{align}
and substitute into the relation of $\partial P_0^2 \text{/} \partial T= 2P_0\partial P_0 \text{/} \partial T$ to give relation between $P_0$ and $\partial P_0^2\text{/}\partial T$.

\begin{figure}[htbp]%h is too strict 
\centering
\includegraphics[width=8.2cm]{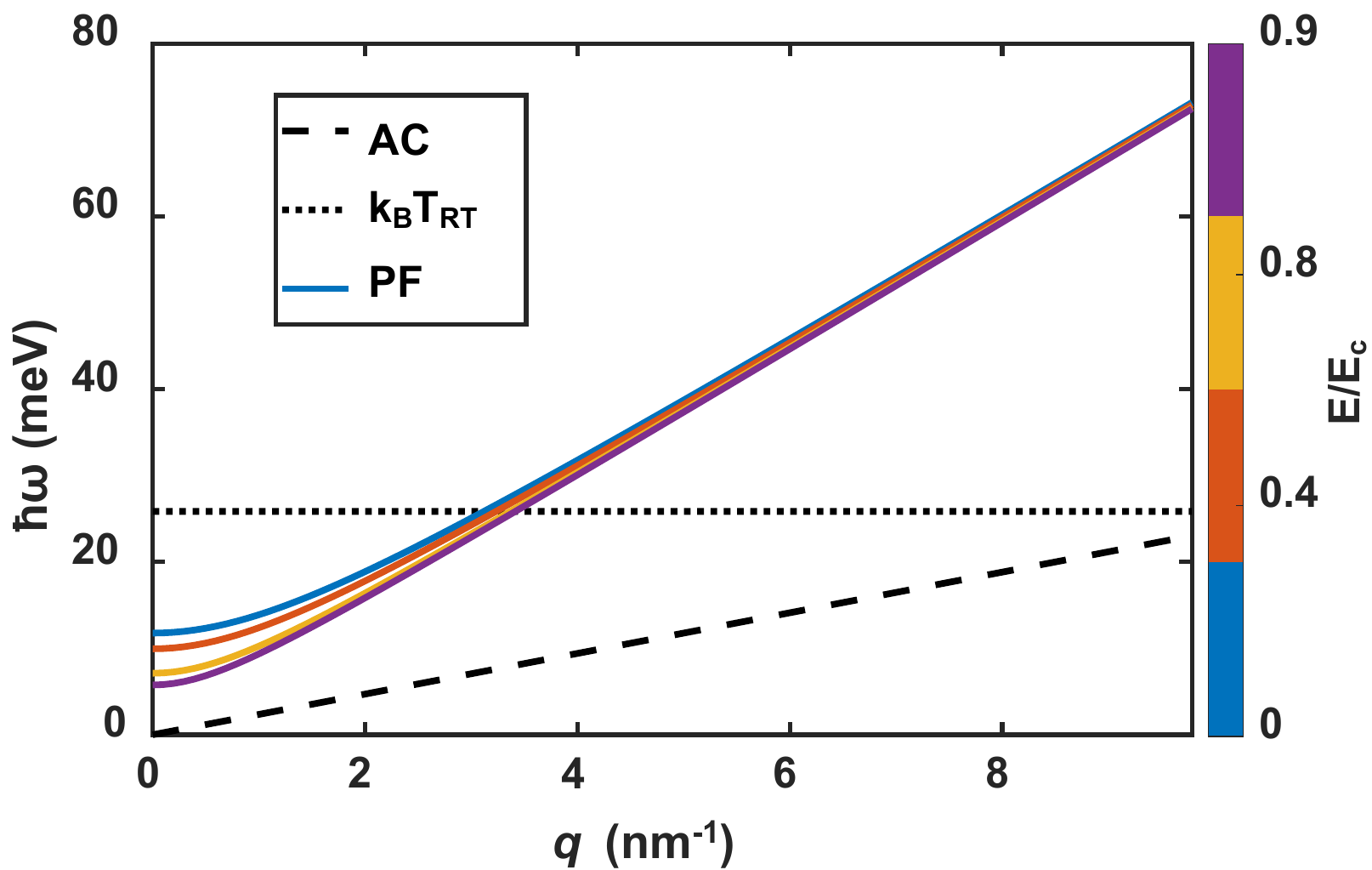}
    \caption{Dispersion relation of ferron in PbTiO$_3$ under different magnitude of electric field oriented oppositely with the spontaneous polarization, compared with the room temperature thermal energy and the corresponding acoustic phonon dispersion relation.}
    \label{fig:figs5}
\end{figure}
As shown in Fig.~\ref{fig:figs5}, when the external electric field oriented in opposite direction to the spontaneous polarization $P_0$, the whole branch of ferron energy spectrum is lowered due to lowered $P_0$.
This leads to a overall lower energy of ferrons and larger population of ferrons, subsequently a enhanced total thermal conductivity.

\section{\label{app:bte}
    Linearized Boltzmann transportation equation with constant relation time approximation}
A temperature gradient perturbs the local equilibrium distribution of elementary excitations, driving the system out of equilibrium. 
Through scattering processes, these excitations relax back towards local equilibrium over a characteristic relaxation time, establishing a steady-state flow of thermal or quasi-particle currents in ferroelectrics. 
We can describe the transport by the semi-classical Boltzmann transportation equation (BTE)~\cite{Callaway2013}:
\begin{align}%\small
    \frac{\partial f}{\partial t}
        +\textbf{v} \cdot \nabla_\textbf{r}f
        +\frac{d\textbf{v}}{dt} \cdot \nabla_\textbf{v}f
    =\left(\frac{\partial f}{\partial t}\right)_s, \label{eqn:bteorg}
\end{align}
where $f(\textbf{r},\textbf{v},t)$ is the distribution of our concerned quasi-particles, and the term $(\partial f/\partial t)_s$ denotes its scattering. 
Under steady state we have $\partial f/\partial t=0$, and under a weak field we ignore the high order third term on the left, then we have only the mutual cancelling diffusion and scattering terms. 
The constant relaxation time approximation (CRTA) gives
\begin{align}%\small
    \left(\frac{\partial f}{\partial t}\right)_s=-\frac{f-\bar{n}_q}{\tau},\label{eqn:btestd}
\end{align}
where $\tau$ is the average relaxation time over $\bm{q}$ and $\bm{v}$. 
Here we have the Boltzmann transport equation with the constant relaxation time approximation (BTE-CRTA):
\begin{align}
    \textbf{v} \cdot \nabla_\textbf{r}f
    =-\frac{f-\bar{n}_q}{\tau}.
\end{align}
Finally, we take the first order linear response $\delta n_q$ of distribution to the weak temperature gradient
\begin{align}%\small
    \delta n_q=f_1-\bar{n}_q=-\tau \frac{d \bar{n}_q}{d T} \bm{v} \cdot \nabla_\textbf{r} T.\label{eqn:btelin}
\end{align}
Therefore, combining the above Eqs. (\ref{eqn:bteorg}-\ref{eqn:btelin}) we have the linearized Boltzmann transport equation with the constant relaxation time approximation (LBTE-CRTA) for the steady state under small temperature field:
\begin{align}%\small
    \textbf{v}\cdot \frac{d \bar{n}_q}{d T}\nabla_\textbf{r} T
    =-\frac{\delta n_q}{\tau}.
\end{align}
The net heat flux density is then
\begin{align}
    \textbf{J}=\frac{1}{V}\sum_{\bm{q}} \hbar \omega_{\bm{q}} \bm{v_q} 
            f(\bm{q}) =\frac{1}{V}\sum_{\bm{q}} \hbar \omega_{\bm{q}} \bm{v_q} 
            \delta n_q(\bm{q}) \nonumber
    ,%\label{eqn:fluxder}
\end{align}
where $f(\bm{q})$ is replaced by $\delta n_q(\bm{q})$ since only the deviating part of distribution from equilibrium contributes to the net flux current.

\section{\label{app:PbTiO$_3$}
    \texorpdfstring{Sound velocity and thermal expansion of $\text{PbTiO}_3$}{Section-3: Sound velocity and thermal expansions of PbTiO3}}

%\subsection{\texorpdfstring{Sound velocity of $\text{PbTiO}_3$}{Sound velocity of PbTiO3}}

\begin{figure}[htbp]%[htbp]%h is too strict 
\centering
\includegraphics[width=7.9cm]{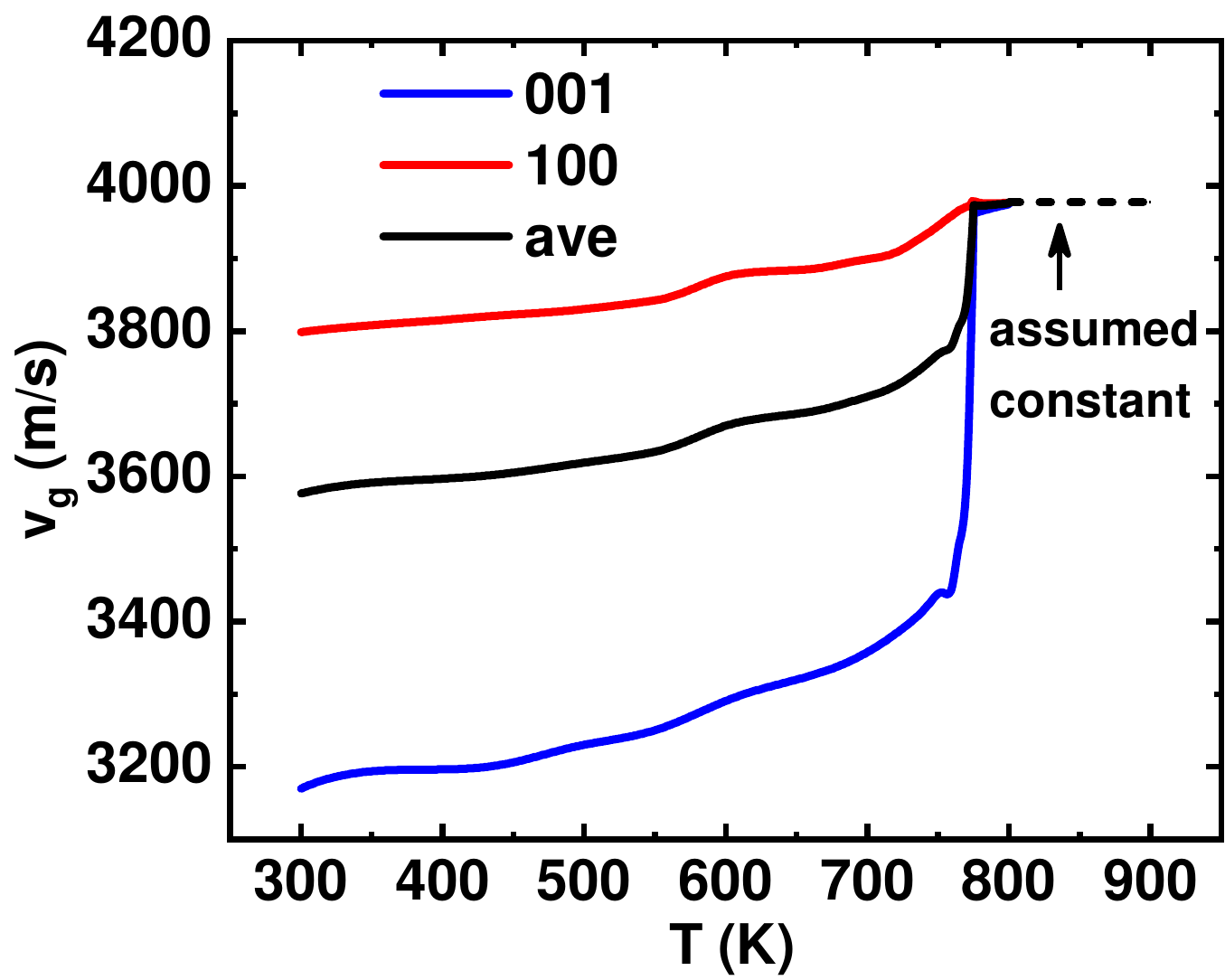}
    \caption{Sound velocity of PbTiO${_3}$ taken from experiment~\cite{Li1996}.
    }
    \label{fig:figs6}
\end{figure}

The sound velocity, or the group velocity of acoustic phonons, is directly extracted and averaged from the experiment of {PbTiO$_3$} single crystal~\cite{Li1996}. 
The average among transverse and longitudinal phonons are taken as
\begin{align}
    v_{\boldsymbol{n}} &= \frac{1}{3} \sum_{\lambda_A}
    \frac{1}{v_{\lambda_A,\boldsymbol{n}}^3},
    ~~~~\boldsymbol{n}= [100]~\text{or}~[001],
\end{align}
with $\lambda_A$ being the branch of the acoustic phonon mode, and the geometrical average among different lattice directions are taken as
\begin{align}
    v_m=\left( v_{[100]}v_{[010]}v_{[001]} \right) ^ {\frac{1}{3}}
       =\left( v_{[100]}^2v_{[001]} \right) ^ {\frac{1}{3}},
\end{align}
as shown in Fig.~\ref{fig:figs6}.

\begin{figure}[b]%h is too strict 
\centering
\includegraphics[width=8.2cm]{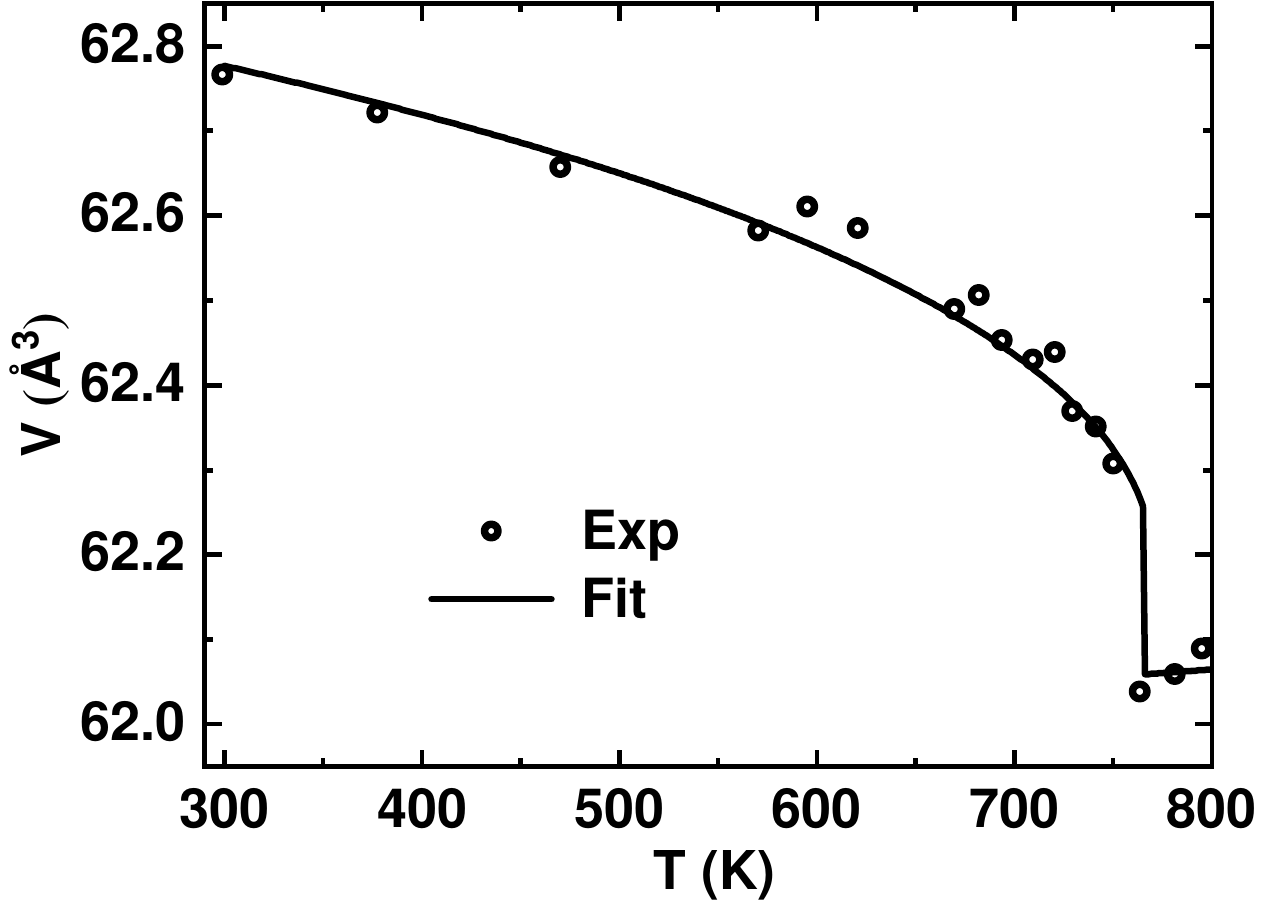}
    \caption{Temperature dependence of PbTiO$_3$ unit cell volume. 
    The experimental results (dots) are from Ref.~\cite{Shirane1951}, and the fitting (line) is based on Eq.~(\ref{eqn:volume}). 
    For simplicity, spontaneous polarization is derived based on the thermodynamic parameters from Ref.~\cite{Haun1987}.
    }
    \label{fig:figs7}
\end{figure}

%\subsection{\texorpdfstring{Thermal expansion of ${PbTiO_3}$}{Thermal expansion of PbTiO3}}
The temperature dependent volume change of ferroelectric PbTiO$_3$ is considered as a superposition of linear thermal expansion and abnormal contraction due to the decrease of polarization~\cite{Shirane1951}. 
The volume change of tetragonal crystal $\Delta V\text{/}V$ is expressed as 
\begin{align}
    \frac{V-V_l}{V_l} 
    &= \epsilon_1+\epsilon_2+\epsilon_3 = (Q_{11}+2Q_{12})P_3^2, \label{eqn:volume1}
\end{align}
where $V_l=V_0+hT$ represents the normal volume expansion with constant expansion coefficient, $\epsilon_i$ are elements of the second-rank strain tensor, $Q_{ij}$ are elements of the fourth-rank electrostrictive coefficient tensor in Voit notation, and $P_j$ are the elements of the first-rank polarization tensor. 
Therefore, the temperature dependent volume of {PbTiO$_3$} is described as
\begin{align}%\small
    V = (V_{\text{ref}} + hT)
    \left[1 + (Q_{11} + 2Q_{12}) P^2 \right], \label{eqn:volume}
\end{align}
with the fitting parameters $V_{\rm ref}=61.93\;\text{\r{A}}^3$, $h=1.731\times10^{-4}\;\text{\AA}^3$/K, and in particular, $(Q_{11}+2Q_{12})=0.02269\;$m$^4$/C$^2$ that is in good agreement with experimentally measured value $0.022\;$m$^4$/C$^2$~\cite{Gavrilyachenko1971,Adachi2001}. 
For simplicity, the values of $P_0 (T)$ used for fitting are derived based on the thermodynamic parameters from Ref.~\cite{Haun1987}, and the fitting results are plotted in Fig.~\ref{fig:figs6}.

%%%%%%%%%%%%%%%%
%\nocite{*} %causes all entries in a bibliography to be printed out
%\bibliography{References}% Produces the bibliography via BibTeX.

\end{document}